\documentclass[12pt]{elsarticle}

\usepackage{amssymb}
\usepackage{amsmath}

\usepackage[utf8]{inputenc}
\usepackage{optidef}
\usepackage{algpseudocode}
\usepackage{mathrsfs}
\usepackage{algorithm}
\usepackage{hyperref}
\usepackage{amsmath,amssymb}
\usepackage{pdflscape}
\usepackage{multirow}
\usepackage{listings}
\usepackage{xcolor}
\usepackage{graphicx}
\usepackage{subcaption}
\usepackage{multicol}
\usepackage[american]{babel}
\usepackage{geometry}
\lstdefinestyle{mypython}{
    language=Python,
    basicstyle=\ttfamily\small,
    keywordstyle=\color{blue}\bfseries,
    stringstyle=\color{red},
    commentstyle=\color{green}\itshape,
    numbers=left,
    numberstyle=\tiny\color{gray},
    stepnumber=1,
    breaklines=true,
    showstringspaces=false,
    frame=single,
    backgroundcolor=\color{gray!10},
}

\journal{Journal of Computational Physics}

\begin{document}
\newcommand{\missing}{THIS HAS TO BE UPDATED}
\newcommand{\algorithmautorefname}{Algorithm}
\newcommand{\myvec}[1]{\boldsymbol{#1}}
\newcommand{\mymatrix}[1]{\boldsymbol{#1}}

\begin{frontmatter}



\title{GPU-accelerated superiorization on constrained physical problems with SupPy} 


\author[label1,label2,label3]{Tobias Becher} 
\author[label4]{Yair Censor}
\author[label4]{Kay Barshad}
\author[label1,label3]{Niklas Wahl}
\affiliation[label1]{organization={Department of Medical Physics in Radiation Oncology, German Cancer Research Center (DKFZ)},
            city={Heidelberg},
            country={Germany}}
\affiliation[label2]{organization={Department of Physics and Astronomy, Heidelberg University},
            city={Heidelberg},
            country={Germany}}
            
\affiliation[label3]{organization={Heidelberg Institute of Radiation Oncology (HIRO) and National Center for Radiation Research in Oncology (NCRO)},
            city={Heidelberg},
            country={Germany}}
            
\affiliation[label4]{organization={Department of Mathematics, University of Haifa},
            city={Haifa},
            country={Israel}}

\begin{abstract}
The superiorization method (SM) is situated between feasibility-seeking and constrained optimization. Instead of aiming at the minimum of a given objective function over a constraint set, it seeks a feasible point at which the objective function value is reduced — though not necessarily minimal — compared to that reached by the feasibility-seeking algorithm alone. This can be advantageous for problems in which the constraints may be inconsistent, in which secondary goals such as noise reduction or regularization describe soft preferences rather than hard targets, or in which a mathematically optimal solution is not strictly required. While the method has been investigated for several applications in physics, its broader use has been limited, in part due to the lack of openly available software for researchers wishing to explore it.

In this work we apply superiorization to three problems from applied physics: seismic image reconstruction, low-dose CT reconstruction and intensity\babelhyphen{hard}modulated radiotherapy treatment planning. These experiments are conducted with SupPy, an open-source modularized Python toolbox developed for this work, 
which supports execution of feasibility-seeking algorithms and their superiorized version on both the CPU and the GPU.
In all three cases the superiorized algorithms achieve favorable results compared to feasibility-seeking alone, with reduced noise in the imaging examples and lowered body dose in the radiotherapy plans. For the radiotherapy case we further observe that superiorization produces clinically viable plans on infeasible constraint sets.

\end{abstract}








\end{frontmatter}

\section{Introduction}

Many problems in physics require solving high-dimensional systems constrained by lower and upper bound conditions. A common approach is to formulate them as constrained optimization problems and apply standard optimization techniques. In some practical applications, however, the optimal solution might not be always required or the consistency of the constraints might not be known in advance, which might lead standard optimization techniques to not converge or fail entirely.

The Superiorization method (SM) uses the same input as a constrained optimization problem, namely, an objective function and a system of constraints, but does not aim at finding an optimal solution. In its most general formulation it can be seen as lying in between feasibility-seeking methods and constrained optimization \cite{Censor2010}. At its base is the use of a projection based feasibility-seeking algorithm (called the basic algorithm) that converges towards a point in the intersection of the (convex) constraints or, in infeasible cases, towards a point of closest proximity to the constraints sets (for appropriate algorithms). The superiorization methodology uses these algorithms and aims to find a point of reduced (i.e., smaller or equal but not necessarily minimal) objective function value compared to a solution of the feasibility-seeking algorithm on its own. For this purpose, a perturbation based on the objective function is performed without disturbing the convergence of the perturbed basic algorithm towards a feasible point.

The method has been investigated for many applications in physics and beyond, see e.g., \cite{Censor2023}.
These include reconstruction of CT images \cite{Sidky2008,Penfold2010,Herman2012, Censor2022b,Nikazad2022a}, radiotherapy treatment planning \cite{Barkmann2023,Gibali2018,Bonacker2020,Borys2025}, magnetoacoustic tomography \cite{Nie2026} or floor planning \cite{Yu2025}, to name a few.
However, many of these approaches are proof-of-principle implementations with specific parameters and few computational effectiveness efforts.
To enable a simpler entry point to the superiorization method for researchers, we propose a modern pythonic approach that flexibly modularizes projection algorithms and perturbation strategies while leveraging both CPU and GPU architectures.

This implementation is tested on three physics application scenarios: The reconstruction of seismic images, CT reconstruction and treatment planning for radiotherapy based on dosimetric constraints. For the reconstruction examples we evaluate different configurations based on runtime, convergence rate and differences to the ground truth. Similarly, the radiotherapy example is evaluated based on runtime, convergence and clinical viability based on dosimetric measures.

We first give a broad overview of the mathematical background of the superiorization method, explain the implementation of the used strategies and then elaborate on the practical physics problems.

\section{Materials and Methods}

\subsection{Mathematical background}
\subsubsection{Projections and feasibility-seeking algorithms} \label{sec:projections}
At the heart of the superiorization methodology are feasibility-seeking
algorithms that aim to find a point in a feasible
set $C:=\cap_{i=1}^{m}C_{i}$, the intersection of a family
of $m$ given constraint sets $\{C_{i}\}_{i=1}^{m}$ in the $n$-dimensional
Euclidean space $\mathbb{R}^{n}$. Such feasibility-seeking algorithms
often use projections onto the sets, where a projection of a point
$\myvec x$ onto a set, say, $C_{i}$ is defined as the point in $C_{i}$
which is closest to $\myvec x$ and denoted by $P_{C_{i}}(\myvec x).$ When the
set is closed and convex then the projection onto the set is guaranteed
to exist and is unique. 
Projections onto a single set $C_{i}$ can be calculated efficiently when
an analytical formula exists; e.g., for a linear constraint $\langle\myvec a^{i},\myvec x\rangle\leq b_{i}$,
the set $C_{i}$ and the formula for projecting onto it are given
by 
\begin{align}
C_{i}= & \{\myvec x\in\mathbb{R}^{n}\mid\langle\myvec a^{i},\myvec x\rangle\leq b_{i}\}\\
P_{C_{i}}(x)= & \begin{cases}
\myvec x-\frac{\langle\myvec a^{i},\myvec x\rangle-b_{i}}{||\myvec a^{i}||^{2}}\myvec a^{i}, & \text{if }\langle\myvec a^{i},\myvec x\rangle>b_{i},\\
\myvec x, & \text{if }\langle\myvec a^{i},\myvec x\rangle\leq b_{i}.
\end{cases}
\end{align}

Adaptations/Relaxations of such projections can be achieved via a
relaxation parameter. For any set $\varOmega\subseteq\mathbb{R}^{n}$
and any real number $\lambda\in[0,2]$ the endpoint of the projection
operator can be modified by letting the outcome, denoted by $P_{\varOmega_{\lambda}}(\myvec x),$
be a point on the line connecting the point $x$ with its projection,
i.e., 

\begin{equation}
P_{\varOmega_{\lambda}}(\myvec x)=\myvec x+\lambda(P_{\varOmega}(\myvec x)-\myvec x).\label{eq:RelaxedProjection}
\end{equation}

Additionally, a step-size function $\sigma(\myvec x)$ can be added
in order to obtain a more general formula: 
\begin{equation}
P_{\varOmega_{\lambda,\sigma}}(\myvec x)=\myvec x+\lambda\sigma(\myvec x)(P_{\varOmega}(\myvec x)-\myvec x)
\end{equation}
with $P_{\varOmega_{\lambda,\sigma}}$ being called an extrapolation
of $P_{\varOmega_{\lambda}}$ when $\sigma(\myvec x)>1$, see \cite[Definition 2.4.1]{Cegielski2013}.

Using these ideas of projecting onto a single set $C_{i}$, feasibility-seeking
projection methods aim to find a point in the intersection of multiple
sets by iteratively projecting onto the individual sets, taking advantage
of the frequently occurring situation in which projections onto the
individual sets are computationally more manageable. To do so, different
methods are used that can broadly be put into four categories: Sequential
algorithms, simultaneous algorithms, block-iterative projection (BIP)
algorithms or string-averaging projection (SAP) methods. Several algorithms
for feasibility-seeking have been published that build and extend
on these principles (\cite{Censor1985,Herman2008,Censor2008,Nikazad2017,Kaczmarz1937,Agmon1954,Motzkin1954}
and many more).

A broad introduction to projection methods and their mathematical
properties is beyond the scope of this work, so we refer the reader
to the vast literature on the topic (e.g., \cite{Cegielski2013,Bauschke1996}); in the sequel we only highlight some of the most important
aspects.

When the intersection of the constraints is nonempty, the above mentioned
methods generate sequences of iterations that converge under reasonable
mathematical conditions, see, e.g., \cite{Butnariu2007,Herman2008a,Davidi2009},
and \cite{Cegielski2025,Nikazad2017}
for some of
their extrapolations. For the simultaneous projection method (and
some of its variants) convergence is even ensured in the case of an
empty intersection. In this case, any iterative sequence generated
by the algorithm will approach a point of closest proximity to the
constraints \cite{Zibetti2018,Nikazad2017,Cegielski2013}. Proximity can be measured by
\begin{equation}
\mathcal{P}(\myvec x)=1/2\sum_{i=1}^{m}w_{i}||P_{C_{i}}(\myvec x)-\myvec x||^{2}.\label{eq:WeightedProximity}
\end{equation}
where $w_{i}$ are weights assigned to the individual constraints,
with $w_i > 0$ and $\sum_{i=1}^{m}{w_{i}}=1$. This function is therefore the "half weighted sum of the squared distances” of a point to the members of the family $\{C_i\}^m_{i=1}$

Another class of problems suitable for feasibility-seeking algorithms
are split problems. These problems consist of an input vector $\myvec x\in C\subseteq\mathbb{R}^{n}$,
an output vector $\myvec y\in Q\subseteq\mathbb{R}^{m}$ and a matrix
$\mymatrix A\in\mathbb{R}^{m\times n}$ connecting the two spaces.
The goal of the ``split-feasibility problem'' then is to find a
vector $\myvec x^{*}$ that fulfills 
\begin{equation}
\myvec x^{*}\in C=\cap_{i=1}^{s}{C_{i}}\text{ such that }\myvec y^{*}=\mymatrix A\myvec x^{*}\in Q=\cap_{j=1}^{t}{Q_{j}},\label{eq:Split}
\end{equation}
see, e.g., \cite{Censor2005}. Among popular approaches
to this problem \cite{Censor2005,Aragon-Artacho2023,Byrne2012},
is the $CQ$-Algorithm of Byrne  \cite{Byrne2012} which
uses separate projection methods onto $C$ and $Q$. Starting from
an arbitrary initialization point $x^{0}\in\mathbb{R}^{n}$ it uses
the iterative process:
\begin{equation}
x^{k+1}=P_{C}(\myvec x^{k}+\gamma\mymatrix A^{T}(P_{Q}(\mymatrix A\myvec x^{k})-\mymatrix A\myvec x^{k})),
\end{equation}
where $\gamma\in(0,2/L)$ with $L$ being the largest eigenvalue of
$\mymatrix A^{T}\mymatrix A$. While this value is computationally
expensive to calculate, in practice an upper bound $L^{*}\geq L$,
using the squared Frobenius norm $\left\Vert \mymatrix A \right\Vert _{F}^{2}$ of
the matrix $\mymatrix A$, can be used instead: \cite{Byrne2009}
\begin{equation}
L^{*}\coloneq\left\Vert \mymatrix A\right\Vert _{F}^{2}=\sum_{i=1}^{m}\sum_{j=1}^{n}{|a_{j}^{i}|^{2}},
\end{equation}
where the $a_{j}^i$'s are the elements of the matrix.

\subsubsection{The superiorization method}\label{subsec:The-superiorization-method}

By defining an objective function $f:\mathbb{R}^{n}\rightarrow\mathbb{R}^ {}$
in addition to the family of $m$ given constraints sets $\{C_{i}\}_{i=1}^{m},$
we have in hand the necessary input for a constrained minimization
problem. However, instead of applying constrained minimization algorithms
we resort to the ``superiorization method''. Extending feasibility-seeking
algorithms, the superiorization method (SM) aims to achieve a reduced
(not necessarily minimal, thus called ``superior'') objective function
value when compared to the solution that the feasibility-seeking algorithm
alone reaches. I.e., if the feasibility-seeking algorithm converges
to point $x^{*}$, the superiorization method aims to generate an
iterative sequence $\{x^{k}\}_{k=0}^{\infty}$ that will converge
to a solution point $x^{**}$ of the feasibility-seeking algorithm
for which $f(x^{**})\leq f(x^{*})$.

This is achieved by perturbing the feasibility-seeking iterates with
negative gradients (or subgradients) of the given objective function
$f$. For a feasibility-seeking algorithm governed by an operator
$T:\mathbb{R}^{n}\rightarrow\mathbb{R}^{n}$ and generating an iterative
feasibility-seeking sequence by the process
\begin{equation}
x^{0}\in\mathbb{R}^{n},\quad \myvec x^{k+1}=\myvec T(\myvec x^{k}),\quad\mathrm{for\;all}\;k\geq0,
\end{equation}
 the perturbed iterative step of the SM can be formulated as 
\begin{equation}
\myvec x^{k+1}=\myvec T(\myvec x^{k}+\beta_{k}\myvec v^{k}),
\end{equation}
where $\beta_{k}$ is a step-size and $\myvec v^{k}$ is a vector
in a direction that leads to a reduced objective function value.

The algorithm is given by the pseudo-code in Algorithm \ref{alg:cap_func_rec}.
The perturbation phase in lines 5 to 18 is followed by the feasibility-seeking
phase.

Feasibility-seeking algorithms whose convergence to a feasible point
is retained in spite of interlacing into them perturbations $\beta_{k}\myvec v^{k}$
which are bounded are called \textit{bounded perturbation resilient}.
Bounded perturbations require that the series of step-sizes does not
diverge $\sum_{k=0}^{\infty}\beta_{k}<\infty$ and that the sequence $(\myvec v^{k})_{k=0}^{\infty}$
is bounded \cite{Censor2010}.

While derivative-free superiorization is possible \cite{Censor2021},
the common way in the literature to calculate the direction vector
$\myvec{v}^{k}$ is using a normalized nonzero subgradient, whenever available. For an objective function $f$ and its subgradient $\myvec g_{f}(\myvec x^{k})$ at point $\myvec{x}^{k}$ it can be expressed as $\myvec{v}^{k}:=\frac{\myvec g_{f}(\myvec x^{k})}{||\myvec{g}_{f}(\myvec{x}^{k})||}$, where the division by the norm is used to keep the sequence bounded.
 To choose the step-size
$\beta_{k}$, a power law $\beta_{k}:=\gamma\alpha^{\ell}$ can be
used. Here, $\gamma$ is a step-size kernel, $\ell$ is an integer that
is related to the iteration index $k$, and is increasing as iterations
proceed, and $\alpha\in(0,1)$ is a step-size parameter to ensure that the step-sizes decrease over time.

While this can substantially reduce the influence of the perturbation
phase in later iterations, Aragón-Artacho et al. \cite{Aragon-Artacho2023}
recently showed that it is possible to ``restart'' the perturbations
to counteract their diminishing effectiveness without interfering
with the bounded resilience of the feasibility-seeking underlying
algorithm. To keep the perturbations bounded, $\ell$ has to increase
after each restart (e.g., by setting $\ell=1$ after the first restart,
$\ell=2$ after the second restart and so on).

Besides using a power law, Nikazad et al. \cite{Nikazad2022a}
introduced a subgradient-projection based method to automatically
choose a step-size compatible with bounded perturbation resilience:

\begin{align}
\beta_{k}:= & \begin{cases}
0, & \text{if } f(\myvec x^{k})\leq f_{\text{ref}}\\
\frac{f(\myvec x^{k})-f_{\text{ref}}}{||\myvec g_{f}(x^{k})||}, & \text{if } f(\myvec x^{k})>f_{\text{ref}},
\end{cases}
\end{align}
with $f_{\text{ref}}$ being a real number that serves as a reference
value, that is automatically updated throughout the iterations and
$g_{f}(x^{k})$ being the (sub)gradient of the objective function
at the current iteration index.

Beyond these perturbation strategies for superiorization, different
approaches like the \textit{Heavy Ball Perturbation} or \textit{Nesterov
perturbations} \cite{Bonacker2021,Bonacker2020}
have been explored, but are beyond the scope of this work.

\begin{algorithm}[hbt!]
\caption{Superiorization: The Superiorized version of a feasibility-seeking
algorithm}
\label{alg:cap_func_rec} \begin{algorithmic}[1] \Require{$\myvec x^{0}$:
Initial point} \Require $\myvec T(\myvec x^{k})$: Feasibility-seeking
algorithmic operator \Require $f$: Objective function \Require $n_{\text{red}}$: number of function
reduction steps in each perturbation phase \State{$k\gets0$} \State{$\myvec x^{k}\gets\myvec x^{0}$}
\While{Stopping criteria not met} \State $n\gets0$ \Comment{Perturbation
phase} \State{$\myvec x^{k,n}\gets\myvec x^{k}$} \While{$n<n_{\text{red}}$}
\Comment{Apply $n_{\text{red}}$ function reduction steps} \State{$loop\gets True$}
\State{$j \gets 0$}
\While{$loop$} \State $\myvec z=\myvec x^{k,n}+\beta_{k,j}\myvec v^{k}$
\State{$j \gets j + 1$}
\Comment{function reduction step} \If{$f(\myvec z)\leq f(\myvec x^{k,n})$}
\State{$n\gets n+1$} \State $\myvec x^{k,n}=\myvec z$ \State{$loop\gets False$}
\EndIf
\EndWhile \EndWhile \State $\myvec x^{k+1}=\myvec T(\myvec x^{k,n})$
\Comment{Feasibility-seeking algorithm iteration} \State $k\gets k+1$
\EndWhile

\end{algorithmic}

\end{algorithm}

\subsection{Implementation}\label{subsec:Implementation}

Each iteration of a superiorization algorithm comprises a perturbation
phase and a feasibility-seeking phase, that modify the current iterate
mostly independently of each other. For general implementations of
the superiorization method it is, therefore, reasonable to decouple
them and allow to set up feasibility-seeking and perturbation strategies
separately. This allows to, e.g., easily explore the effect of different
perturbation strategies by keeping the corresponding feasibility-seeking
algorithm fixed. We followed this approach while constructing \texttt{SupPy} \cite{Becher2026},
an open-source $Python$ native toolbox for superiorization that we
built for this work. \footnote{
The \texttt{SupPy} toolbox is available on \href{https://github.com/DKFZ-OpenMedPhys/SupPy}{GitHub} with a \textit{BSD 3-Clause License}.  Scripts used to generate and run the algorithms are available on a dedicated \href{https://github.com/DKFZ-OpenMedPhys/SupPy/tree/research/paper-suppy}{branch} based on \textit{v0.4.0}.} 
The modular structure that we adopted allows different feasibility-seeking
algorithms and different perturbation strategies to be implemented
as classes in respective modules and can be combined through classes
from the overarching superiorization module.

A more in-depth explanation of these main components are further explained
in the following.

\subsubsection{Feasibility-seeking}\label{subsec:Feasibility-seeking}

Feasibility-seeking algorithms can cover many types of problems, from
small scale examples intended for the study and verification of algorithms,
to large scale linear problems that thrive with dedicated implementations.
Both of these are possible within \texttt{SupPy}.

Smaller problems can be set up by instantiating individual constraints
and treating them with an appropriate projection method (e.g., a sequential
or a simultaneous projections method).

A simple example of how such a problem can be set up in \texttt{SupPy} is
presented in \autoref{lst:Test}.

\begin{lstlisting}[style=mypython,label = lst:Test,caption={Example of how to set up a small scale feasibility-seeking example in \texttt{SupPy}. The goal of the problem here is to find the intersection of two balls via a sequential projection.},captionpos=b]
import numpy as np
from suppy.projections import BallProjection,SequentialProjection
center_1,center_2 = np.array([1,1]),np.array([-0.5,-1])
radius_1,radius_2 = 2,1
ball_1 = BallProjection(center_1,radius_1)
ball_2 = BallProjection(center_2,radius_2)
joined_projection = SequentialProjection([ball_1,ball_2])

\end{lstlisting}

While this approach could also be used for larger problems, it becomes
computationally inefficient and faster runtimes can be achieved by
employing matrices, as done in several dedicated implementations in
\texttt{SupPy}.

Apart from these common problem formulations, projections can be fully
customized by deriving them from a $CustomProjection$ interface of
\texttt{SupPy}.

Projection based feasibility-seeking algorithms all aim to find a
point inside a feasible set or generate an iterative sequence that
converges towards a point of closest proximity to all individual sets,
see Subsection \ref{sec:projections}. To provide
a measure of distance from the feasible set or check whether a point
is feasible, each algorithm needs to evaluate the proximity of a current
iterate to the constraints sets.

To account for all of this, each algorithm requires a method to calculate
the updated iteration of the current iteration and a proximity measure.
Projection algorithms further require a ``control sequence'' that
dictates in which order update steps are performed until feasibility
is reached or a suitable stopping criterion is met. In \texttt{SupPy} this
is realized through a $BaseProjection$ class that provides an interface
for the $project$ and $proximity$ functions, which all projections
and projection algorithms are using.

Proximity for individual constraints can be calculated through analytical
formulas, while algorithms combining several constraints evaluate
a weighted sum of the distances to the comprising constraints, see
Equation \ref{eq:WeightedProximity}.

For proximity, \texttt{SupPy} implements an extension of the weighted sum
of squared distances. I.e., a $p$-norm like general power sum with
the power $p$, chosen by the user (with default $p=2$). Denoting
the distance to the $i$-th constraint by $\mathcal{P}_{i}$ we use,

\begin{equation}
\mathcal{P}:=\sum_{i=1}^{m}\myvec w_{i}{(\mathcal{P}_{i})^{p}}.
\end{equation}

A max norm measure is also available, 
\begin{equation}
\mathcal{P}_{\max}:=\max_{1\leq i\leq m}{\mathcal{P}_{i}}.
\end{equation}

For split-feasibility problems, proximity measures are evaluated independently
for the input and the target space. Based on these notions, the standard
stopping criterion for each algorithm checks the following three conditions:
\begin{enumerate}
\item Whether the proximity value $\mathcal{P}$ drops below a pre-defined
threshold $\mathcal{P}^{*}$ (which is by default $10^{-6}$): $\mathcal{P}\leq\mathcal{P}^{*}.$
\item Whether the relative change in proximity value $\Delta\mathcal{P}:=|\frac{\mathcal{P}^{k+1}-\mathcal{P}^{k}}{\max\{1,\mathcal{P}^{k}\}}|$
, where $\mathcal{P}^{k}$ is the value of $\mathcal{P}^ {}$ at the
$k$-th iteration, drops below a threshold $\Delta\mathcal{P}^{*}$
for $N$ consecutive iterations (by default $\Delta\mathcal{P}^{*}=10^{-8}$,
$N=5$).
\item Whether a maximum number of iterations $N_{\max}$ is reached (which
is by default $N_{\max}=500$).
\end{enumerate}
These criteria ensure that the algorithm terminates either when the
current iterate is feasible or very close to feasibility, or when
only small changes of the proximity value can be achieved by additional
iterations, or when a maximal number of iterations is reached. For
split-feasibility-seeking algorithms these criteria are only evaluated
for the ``target space'' $\mathbb{R}^{m},$ consult (\ref{eq:Split}).
Individual user-defined stopping criteria can be added through a callback
interface.

While feasibility-seeking algorithms can be evaluated on their own
they can also be combined with a perturbation strategy for a superiorized
version.

\subsubsection{Perturbations}\label{subsec:Perturbations}

Perturbation strategies present an option to improve upon the current
iterates of a feasibility-seeking algorithm with respect to some objective
function. Commonly, this is achieved through a gradient step, although
derivative free updates are also possible, see, e.g., \cite{Censor2019}. To
choose the step-size for gradient based strategies, a common approach
in the literature is to use a decreasing power law, although adaptive
step-sizes based on subgradient directions \cite{Nikazad2022a} have been shown
to work, too. Both of these methods are implemented as individual
classes in \texttt{SupPy}.

To initialize the perturbations, function handles for the objective
function $f$ and its (sub-)gradient $\myvec g_{f}$ are required,
while additional parameters like the kernel $\gamma$ for the power
law, the parameter $\alpha,$ or the number of function reduction
steps are treated as optional parameters.

While derivative free or higher order perturbation strategies are
currently not incorporated in \texttt{SupPy}, the current class based modular
structure would allow easy future extensions.

\subsubsection{Superiorization}\label{subsec:Superiorization}

Combination of a feasibility-seeking algorithm with a perturbation
strategy is achieved in \texttt{SupPy} through the ``superiorization module''.
Creation of a superiorization algorithm requires an instance
of a feasibility-seeking algorithm and a compatible perturbation strategy. In case of applicable split-feasibility-seeking problems,
perturbation strategies can be applied separately to the input $\mathbb{R}^{n}$
and the output $\mathbb{R}^{m}$ spaces (see \cite{Aragon-Artacho2023}).
The main $solve$ function then runs the algorithm as defined in \autoref {alg:cap_func_rec}
until a suitable stopping criterion is met. \autoref{lst:superiorization}
shows an example of how the setup and execution of an algorithm may
look like.

Considering the numbered list of stopping conditions in Subsection \ref{subsec:Feasibility-seeking},
for the default termination criterion, individual checks for the feasibility-seeking
algorithm and the objective function have to be met.

For the former, either criterion (1) or criterion (2) for the feasibility-seeking
algorithms introduced in the previous section need to be met. 

For the latter, a criterion similar to criterion (2) is employed,
requiring that the relative change in objective function value $\Delta f:=\left|\frac{f^{k+1}-f^{k}}{\max\{1,f^{k}\}}\right|,$
where $f^{k}:=f(x^{k}),$ 
falls below a threshold $\Delta f^{*}\leq10^{-6}$. If the objective
function criterion is met, as well as one of the criteria on the feasibility-seeking
algorithm, the procedure stops. Otherwise, it continues until a maximum
number of iterations is reached.

\begin{lstlisting}[float,floatplacement=H,style=mypython,label = lst:superiorization,caption={Example of how to set up a small scale superiorization example in \texttt{SupPy}. The goal of the problem here is to find the intersection of two balls via a sequential projection with an additional objective to reduce the distance to the origin.},captionpos=b]]
import numpy as np
from suppy.projections import BallProjection,SequentialProjection
from suppy.perturbations import PowerSeriesGradientPerturbation
from suppy.superiorization import Superiorization

func_1 = lambda x: x@x
grad_1 = lambda x: 2*x

#set up model
center_1, center_2 = np.array([1.2,0]), np.array([0,1.4])
radius = 1
ball_1 = BallProjection(center_1, radius)
ball_2 = BallProjection(center_2, radius)
x0 = np.array([2.5,1.5])

proj = SequentialProjection([ball_1,ball_2])

pert = PowerSeriesGradientPerturbation(func_1,grad_1)

sup = Superiorization(proj,pert)

#solve the problem
x_proj = proj.solve(np.array([2.5,1.5]),storage =True)

xF = sup.solve(np.array([2.5,1.5]),10,storage = True)
\end{lstlisting}

\subsubsection{Interoperability}\label{subsec:Interoperability}

Many feasibility-seeking algorithms and perturbation strategies rely
on large scale matrix operations that can be executed faster on a
GPU than on a CPU. In other cases however, e.g., large scale sequential
algorithms, a CPU evaluation might be faster.

To enable both options, all implementations in \texttt{SupPy} allow the usage
of the CPU as well as the GPU. This is achieved by using the \texttt{NumPy}
and \texttt{SciPy.sparse} libraries for CPU and the \texttt{CuPy} library for
GPU operations. \texttt{SupPy} will, for the most part, automatically detect
whether an algorithm should be executed on the CPU or on the GPU based
on the input types and then act accordingly.

\subsection{Applications}\label{subsec:Applications}

To verify the usability of the toolbox we present and analyze different
applications from the fields of seismic wave tomography, CT reconstruction
and treatment planning for radiotherapy. All scripts were run on
a Lenovo ThinkPad P16 Gen2 notebook with a \textit{$13^{th}$ Gen
Intel(R) Core(TM) i9-13980HX 2.20 GHz, 64\,GB RAM CPU} and an \textit{NVIDIA
RTX 3500 Ada Generation Laptop GPU}. In the next subsections, different
algorithms that are available in the toolbox are highlighted through
applications in the realm of physics.

\subsubsection{Seismic Tomography}
The first problem we showcase is in the field of seismographic tomography.
Here, spatial information about the medium traversed by seismic waves
is reconstructed from propagation time measured between source-receiver
pairs. The reconstruction problem can be modeled as an approximate
system of linear equations, 
\begin{equation}
\mymatrix A\myvec x\approx\myvec b\label{eq:seismic}
\end{equation}
where $\myvec b$ is the measured (noisy) data, $\myvec x$ is the
(discretized) $N\times N$ structure, reformulated as a single column
vector, and $\mymatrix A$ is a matrix encoding the physics of the
reconstruction problem. A solution of these equations can be achieved
through least-squares minimization problem ($\min||\mymatrix A\myvec x-\myvec b||_{2}^{2}$),
with possible additional regularization to counteract the noise and
steer the reconstruction towards acceptable physical solutions \cite{Loris2010,Charlety2013}.

We explored this problem with \texttt{SupPy} based on the reconstruction
of $2$-dimensional (2D) tectonic plates. In contrast to the least-squares
minimization approach, a projection based feasibility-seeking algorithm
was used to solve the system of linear equations while different perturbation
strategies were applied, to perform a superiorization.

The underlying problem was generated using the \textit{seismicwavetomo}
setup of the \texttt{AIR Tools II Toolbox} \cite{Hansen2018},
which uses a simplified model that assumes that waves travel only
in the first Fresnel zone. 64 sources and 128 receivers were used
to simulate the measurements on an underlying structure of $256\times256$
pixels, leading to a matrix $\mymatrix A$ of size $8192\times65536$.
Since the problem $\mymatrix A\myvec x=\myvec b$ has an exact solution,
we also investigated a modified version with $5\%$ Gaussian noise
added to the ``measurements'' vector $\myvec b$ in order to simulate
a more realistic setting.

For feasibility-seeking we used the Diagonally Relaxed Orthogonal Projection
(DROP) algorithm \cite{Censor2008} which is also present
in the CPU-based \texttt{AIR Tools II Toolbox} which implements several
sequential and simultaneous reconstruction techniques in Matlab. The
DROP algorithm is a variation of the simultaneous projection that
takes the sparsity pattern of matrix $\mymatrix A$ into account (see
the Appendix at the end of the paper).

Besides the feasibility-seeking on its own, we also investigated the
superiorized algorithm by evaluating three different perturbation
approaches on this problem. In particular, we looked at different regularization
techniques ($L1$, $L2$ and Total Variation ($TV$)-regularization)
and performed perturbation strategies inspired by them.

Total Variation for a 2D $N\times N$ image $\myvec X$ can be computed
as 
\begin{equation}
TV(\myvec X)=\sum_{i=1}^{N-1}\sum_{j=1}^{N-1}{\sqrt{(\myvec X_{i+1,j}-\myvec X_{i,j})^{2}+(\myvec X_{i,j+1}-\myvec X_{i,j})^{2}}}
\end{equation}
where $\myvec X_{i,j}$ is the image brightness at the $(i,j)$-th
pixel, see, e.g., \cite{Penfold2010}, and relies on the assumption
that real world images usually do not contain random noise, but contain
gradual changes. To calculate the subgradient of the $TV$, required
for the perturbations, we used in this work the methodology explained
by Combettes et al. \cite{Combettes2002,Penfold2010}.

All perturbations were performed using the standard perturbation strategy
described in Subsection \ref{subsec:The-superiorization-method}.
In particular, the step-sizes $\beta_{k}$ were updated here according
to a power law $\beta_{k}=\gamma\alpha^{\ell}$ with $\gamma=1$ and
$\alpha=0.975$, where $\ell=\ell(k)$ is an integer that depends
on the iteration index $k$. Four reduction steps were performed in
each perturbation phase and $\ell$ was restarted after every 50 iterations,
see \cite{Aragon-Artacho2023}.

The perturbations were only performed in \texttt{SupPy} as \texttt{AIR Tools
II} is not intended for such modifications.

To quantify how good an iterate $\myvec x^{k}$ approximates the true
structure $\myvec x_{T}$, the relative error was used, calculated
via 
\begin{equation}
\epsilon=\frac{||\myvec x^{k}-\myvec x_{T}||}{||\myvec x_{T}||}.\label{eq:rel_error}
\end{equation}

\subsubsection{CT image reconstruction}
Reconstruction of Computed Tomography (CT) images has been at the
forefront of applications of superiorization \cite{Sidky2008,Penfold2010,Herman2012,Censor2022b,Nikazad2022a}.
It is a computationally similar, but physically different, concept
to that of seismic image reconstruction discussed above. For CT reconstruction,
x-ray induced line integrals are taken from various angles, with each
direction encoding different information of the patient's anatomy.
Well-known methods to retrieve the anatomical information from these
line integrals are the filtered back-projection (FBP), the class of
iterative reconstruction techniques, as well as more recent deep-learning
based approaches \cite{Hsieh2013,Szczykutowicz2022}.
The problem to solve is a linear system of equations, similar to \autoref{eq:seismic},
with $\mymatrix A$ in this case encoding the information on how much
each pixel is contributing to the measurements $b_{i}$ of the $i$-th
line integral.

Superiorization for CT reconstruction with \texttt{SupPy} is demonstrated
on the Benchmark Dataset for Low-Dose CT (LoDoPaB-CT) dataset \cite{Leuschner2021}.
It consists of $362\times362$ CT images and corresponding sinograms
based on simulated low dose-measurements acquired with scans taken
over 1000 equidistant angles between 0° and 180°, each measurement
being recorded by 513 evenly-spaced detector bins, resulting in a
$(513,000)\times(131,044)$ sparse reconstruction matrix $\mymatrix A$.
While the dataset does not
offer the reconstruction matrix $\mymatrix A$ directly, it can be extracted from the ASTRA-toolbox \cite{vanAarle2015,Aarle2016a}
used to generate the dataset.

We present the reconstruction process using different feasibility-seeking
algorithms and different perturbation strategies of the superiorization
algorithm that are available in \texttt{SupPy}. The first experiment looked
at different feasibility-seeking algorithms, evaluating their runtime
and convergence behavior on their own, as well as in combination with
a fixed perturbation strategy of the superiorization algorithm. A
second experiment compared the effect of different perturbation strategies
of the superiorization algorithm evaluated on the same feasibility-seeking
algorithm.

In the first experimental setup, we compared the following sequential
and simultaneous feasibility-seeking algorithms.
\begin{enumerate}
\item The sequential projection method of Kaczmarz algorithm \cite{Kaczmarz1937},
also known in the field of image reconstruction as the Algebraic Reconstruction
Technique (ART) see, e.g. \cite{Herman2009}.
\item The Extrapolated Landweber (EL) method \cite{Cegielski2013,Cegielski2025}.
\item The Conjugate Gradient (CG) method \cite{Zibetti2018}.
\item A modification of the Landweber algorithm with built-in error minimization
relaxation (EMR) \cite{Nikazad2017}.
\end{enumerate}
For the simultaneous algorithms runtimes were evaluated for the CPU
and GPU, while the sequential algorithm was only run on the CPU. An
overview of the different update steps used can be found in the Appendix
below.

For the perturbation in the first experiment, a kernel $\gamma=5$,
step size parameter $\alpha=0.99$ and $n_{\text{red}}=4$ function reduction
steps in each perturbation phase were used. In addition, restarts \cite{Aragon-Artacho2023} were
performed every $n_{restart} = 50$ iterations.

In the second experimental setup different perturbation strategies
in combination with the EMR algorithm were tested. In particular,
variations of parameters for the power series perturbation were explored,
as well as an adaptive perturbation strategy. A summary of the parameters
used for the power series is found in \autoref{tab:perturbation_parameters}.
For the adaptive perturbation strategy no step-size parameters had
to be chosen and the initialization was performed as described in
\cite{Nikazad2022a}.

\begin{table}[ht]
    \centering
        \begin{tabular}{lccccc}
         & $\alpha$ & $\gamma$ & $n_{\text{red}}$ & $n_{restart}$ \\
        \hline
         Strategy 1 & 0.99 & 5 & 4 & 50 \\
         Strategy 2 & 0.99 & 5 & 4 & $\diagup$ \\
         Strategy 3 & 0.9 & 5 & 4 & 50 \\
         Strategy 4 & 0.9 & 5 & 4 & $\diagup$ \\
         Strategy 5 & 0.9 & 1 & 1 & $\diagup$ \\
        \hline
        Strategy 6 & \multicolumn{4}{c}{Adaptive}
        \end{tabular}
    \caption[Par] {Parameters used for the power series perturbation strategies. For the initialization of the adaptive strategy see \cite{Nikazad2022a}.}
    \label{tab:perturbation_parameters}
\end{table}

Besides running for a fixed number of iterations and analyzing the performance of the algorithms, we also looked at a custom stopping criterion based on a variance measure of consecutive iterates \cite{Nikazad2022a}.
In each iteration the variance of the relative change between iterates was evaluated
\begin{align}
    \mathrm{Var}(\mathbf{\myvec{\omega}^k}) & = \frac{1}{k-1} \sum_{i=1}^{k} (\omega^k_i - \overline{\myvec\omega}^{k})^2 \\
    \text{with } \myvec\omega^k&  = \left(
    \frac{||\myvec{x}^1 - \myvec{x}^0 ||}{||\myvec{x}^1||}, ...., \frac{||\myvec{x}^{i+1} - x^{i} ||}{||\myvec{x}^{i+1}||}, ...,\frac{||\myvec{x}^{k} - \myvec{x}^{k-1}}{||\myvec{x}^{k}||}\right),
\end{align}
where $\overline{\myvec\omega}^{k}$ denotes the mean of the vector at iteration $k$.
Once $\mathrm{Var}(\myvec{\omega}^k)$  falls below a predetermined threshold (here 0.01) the algorithm stops.
This can be used to stop the reconstruction process early and therefore counteracts the overfitting of the noise in later iterations which leads to an increase in relative error (see \autoref{eq:rel_error}).

\subsubsection{Radiotherapy treatment planning}
The next application we highlight is in the field of treatment planning for radiotherapy. Radiotherapy is a modality used for treating cancer, with around $50\%$ of all cancer patients receiving it during their treatment \cite{Lievens2020}. For each individual patient a suitable setup of the radiation field delivered by the radiation machine has to be found based on their anatomy and on the medical prescription. Arriving at such a treatment plan involves several steps, starting with the acquisition of a CT of the patient. Based on this image, structures of the patient, like the tumor volume and surrounding organs, are segmented.
The next step then deals with finding a suitable irradiation setup through modeling the situation as a feasibility-seeking problem or as an optimization problem.

While several parameters such as the beam angles can be varied, the main mathematical feasibility or optimization problem in treatment planning deals with finding suitable intensities of the beams $\myvec w$ that allow delivery of a specific dose distribution $\myvec d$.

For this a discretized model is used: The 3D CT image of the patient is divided into small cubic volumes called $voxels$ (which are analogous to pixels in a 2D picture) while the external radiation beams are separated into smaller subelements called beamlets with variable intensity $w_j$ of each. The dose delivered to the voxels can be expressed as a vector $\myvec{d}  = (d_{i})_{i=1}^{m} \in \mathbb{R}^m$ in the "$dose$-space", while the intensities can be modeled as the vector $\myvec w =(w_{j})_{j=1}^{n}\in \mathbb{R}^n$ in the "$intensity$-space").
The physics of the model is condensed into the $m\times n$ (pre-computed) dose influence matrix $\mymatrix{A}=(a_{ij})_{i=1,j=1}^{m,n}$. Each element $a_{ij}$ in $\mymatrix{A}$ is the dose absorbed in voxel $i$ due to a unit of intensity along the $j$-th beamlet. This means that 
\begin{equation}
    d_i = \sum_{j=1}^n a_{ij} w_j
\end{equation}
is the total dose absorbed in voxel $i$ due to an intensity vector $\myvec{w}$.

To find a suitable dose distribution coverage of the tumor has to be weighed against sparing of surrounding organs at risk. For this, goals on the different organs, as defined by a clinician, are translated into mathematical objective functions and constraints. For example the dose in the tumor volume may have to be kept between lower and upper bounds, $d_{\min}$ and $d_{\max}$, respectively. This is modeled as a constraint 
\begin{align}
    d_{\min} \leq d_i \leq d_{\max} & \quad \quad \forall i \in S,
\end{align}
for all $i\in S$, where $S$ is the set of voxels that belong to the structure of interest.

A different goal of reducing the mean dose in an organ at risk surrounding the tumor, could be modeled as an objective function that should be minimized 
\begin{align}
    f_{Mean}(\myvec{d}) = \frac{1}{N_S}\sum_{i \in S} d_i
\end{align}
where $N_{S}$ is the number of voxels in the structure $S$. 

Further options are dose-volume constraints (DVC) that come in two different variants:
\begin{align}
    D_{V_{\%}} \leq d_{ref} \\
    D_{V_{\%}} \geq d_{ref}
\end{align}
Here $D_{V_{\%}}$ stands for the dose absorbed in voxels that constitute a fraction of $V_{\%}$ of all voxels included in a certain organ, without regard to which specific voxels are the ones that absorb this. The first means  that not more than $V_{\%}$ of the voxels of an organ should absorb a dose higher than $d_{ref}$. The second requires the opposite - that a specific dose $d_{ref}$ needs to be delivered to at least $V_{\%}$ of the voxels of an organ.

Combining all such objectives and constraints can be formulated as the following constrained optimization problem.
\begin{mini}
    {\myvec{w}}{\myvec{f}(\myvec{d}) = \sum_l p_\ell f_\ell(\myvec{d})
}
    {\label{eq:dose_opti}}{}
    \addConstraint{\myvec{d} \in C}{ = \cap_k C_k }{}
    \addConstraint{\myvec{d}=}{\mymatrix{A}\myvec{w}}{}
    \addConstraint{\myvec{w}\geq}{ 0}{}
\end{mini}

Here $f_{\ell}$ are objective functions (with associated penalties $p_{\ell}$), $C_{k}$ are constraints sets and $\myvec w\geq0$ a physical constraint since beam intensities cannot be negative.

If there are no objective functions $f_{\ell}$ involved then solving to find a vector $\myvec{w}$ that will obey the constraints is a feasibility-seeking problem.

An alternative approach to solving the full-fledged constrained optimization problem (\autoref{eq:dose_opti}) is to use these data in a superiorization approach.

This is explored here in the context of conflicting constraints, that lead to infeasibility, with an underlying split-feasibility-seeking algorithm (see \autoref{eq:Split}).
Of particular interest is it here, since if $C=\emptyset$ is the empty set then a constrained optimization algorithm cannot work at all.

We demonstrate this on a simple structure with a horseshoe-shaped tumor wrapped around a circular organ at risk, as well as by a clinical patient case. Both cases are part of the \texttt{CORT} dataset \cite{Craft2014a} and openly available. 

Each setup consisted of 9 photon beams arranged equally around the patient. The dose-influence matrix $\mymatrix{A}$ was calculated using the open-source radiotherapy treatment planning software \texttt{matRad} \cite{Ackermann2020,Wieser2017}.  Each patient had several constraints assigned to various structures and a further objective function to reduce the dose in the surrounding body (see \autoref{tab:final_TG119_plan} and \autoref{tab:final_HN_plan}, respectively). The constraints in both cases were constructed in such a way, that they cannot be fulfilled simultaneously, leading to the problem being infeasible, which makes it non applicable for many common constrained optimization solvers. To demonstrate this for the horseshoe phantom, the same set of objectives and constraints was passed to the Interior Point OPTimizer \texttt{IPOPT} solver \cite{Wachter2006}. For the head-and-neck plan a reference plan was calculated on optimizable objectives and constraints to assess the clinical relevance.

The constraints employed in the two setups comprised both linear constraints and dose-volume constraints (DVCs). These were modeled as a split-feasibility-seeking problem (\autoref{eq:Split}) with the linear constraints modeled as the constraints $C_k$ and the DVC constraints modeled as the constraints $Q_j$. For projections onto the $C_k$ constraints a modification of the EMR iterate suitable for linear inequalities was applied, while for the DVC constraints the methodology described by Penfold et al.\cite{Penfold2017} was used. Lastly, negative weights were set to zero after each iteration to maintain the non-negativity requirement $\myvec{w} \geq 0$, which was performed through a \textit{BoxProjection} onto the nonnegative orthant $\mathbb{R}^n_{\text{+}}$ in \texttt{SupPy}.

Perturbations for superiorization were performed with the step sizes chosen by a power law $\beta_k = \gamma \alpha^\ell$ with $\gamma=1,\alpha=0.7$ for the horseshoe-shaped phantom and $\gamma=1,\alpha=0.99$ for the head and neck patient. Both algorithms performed restarts after each consecutive 500 iterations.

\begin{table}[ht]
    \centering
    \begin{tabular}{lll}
    VOI/Structure &  Constraints & Objective\\
    \hline
    \multirow{3}{*}{Target}& $D_{95\%} \geq 2\,Gy$ &\\
    & $D_{5\%} \leq 2.17\,Gy$& \\
    & $1.93\,Gy \leq d \leq 2.27\,Gy$\\
    \cline{1-2}
    Core & $D_{5\%} \leq 0.33\,Gy$ & \\
    \cline{1-2}
    Body & & $f_{Mean}$ \\
    \end{tabular}
    \caption[Dose objective functions] {Constraints and objectives used for the horseshoe-shaped tumor plan calculation. Superiorization uses all components, while the feasibility-seeking algorithm considers only the constraints.}
    \label{tab:final_TG119_plan}
\end{table}

\begin{table}[ht]
    \centering
    \resizebox{\textwidth}{!}{%
    \begin{tabular}{l|ll|ll}
    & \multicolumn{2}{c|}{Superiorization} & \multicolumn{2}{c}{Optimization}\\
    VOI/Structure &  Constr. & Obj. & Constr. & Obj.\\
    \hline
    Skin & $ \myvec{d} \leq 2.17\,Gy$ & $f_{Mean}$ &$ \myvec{d} \leq 2.17\,Gy$ & $f_{Mean}$\\
    \cline{1-2}

    PTV$_{2.1}$ & $2.03\,Gy \leq \myvec{d} \leq 2.17\,Gy$ && &$f_{SqDev}(2.1\,Gy)$\\
    \cline{1-2}

    PTV$_{2.33}$ & $2.27\,Gy \leq \myvec{d} \leq 2.40\,Gy$ && &$f_{SqDev}(2.33\,Gy)$\\
    \cline{1-2}   

    \multirow{2}{*}{Brain stem PRV}& $D_{5\%} \leq0.33\,Gy$ && &\multirow{2}{*}{$f_{SqOD}(0.33\,Gy)$}\\
    & $ \myvec{d} \leq 1\,Gy$ & &&\\
    \cline{1-2}
    
    \multirow{2}{*}{Spinal cord PRV}& $D_{5\%} \leq0.33\,Gy$  && &\multirow{2}{*}{$f_{SqOD}(0.33\,Gy)$}\\
    & $ \myvec{d} \leq 1.33\,Gy$& &&\\
    \cline{1-2}

    \multirow{2}{*}{Parotid LT}& $D_{5\%} \leq0.33\,Gy$ & && \multirow{2}{*}{$f_{SqOD}(0.33\,Gy)$}\\
    & $ \myvec{d} \leq 1.33\,Gy$& &&\\
    \cline{1-2}

    \multirow{2}{*}{Parotid RT}& $D_{5\%} \leq0.33\,Gy$ & &&\multirow{2}{*}{$f_{SqOD}(0.33\,Gy)$}\\
    & $ \myvec{d} \leq 1.33\,Gy$& & \\
    \cline{1-2}
    
    \multirow{2}{*}{Larynx}& $D_{5\%} \leq0.33\,Gy$ & & &\multirow{2}{*}{$f_{SqOD}(0.33\,Gy)$}\\
    & $ \myvec{d} \leq 1.33\,Gy$ & &&\\
    \cline{1-2}
    
    \end{tabular}
    }
    \caption[Dose objective functions] {Constraints and objectives used for the head and neck plan calculation. Superiorization uses all components, while the feasibility-seeking algorithm considers only the constraints. For the optimized plan a \textit{squared} \textit{deviation} objective  function is used instead of the min-max dose constraints on the target and for most organs at risk a \textit{squared} {overdosing} objective function is used. A further explanation with formulas for the objective functions can be found in the appendix (\autoref{tab:dose_objectives}).}
    \label{tab:final_HN_plan}
\end{table}

\newpage

\section{Results}

\subsection{Application 1: Seismic image reconstruction}

Reconstructions with DROP in \texttt{SupPy} and \textit{AIR Tools II} on the same machine resulted in numerically identical solutions. However, running 1000 iterations of DROP took roughly $100\,s$ with \textit{AIR Tools II}, whereas the GPU-accelerated evaluation in \texttt{SupPy} took less than $10\,s$.

\begin{figure}
    \centering
    \includegraphics[width = \textwidth]{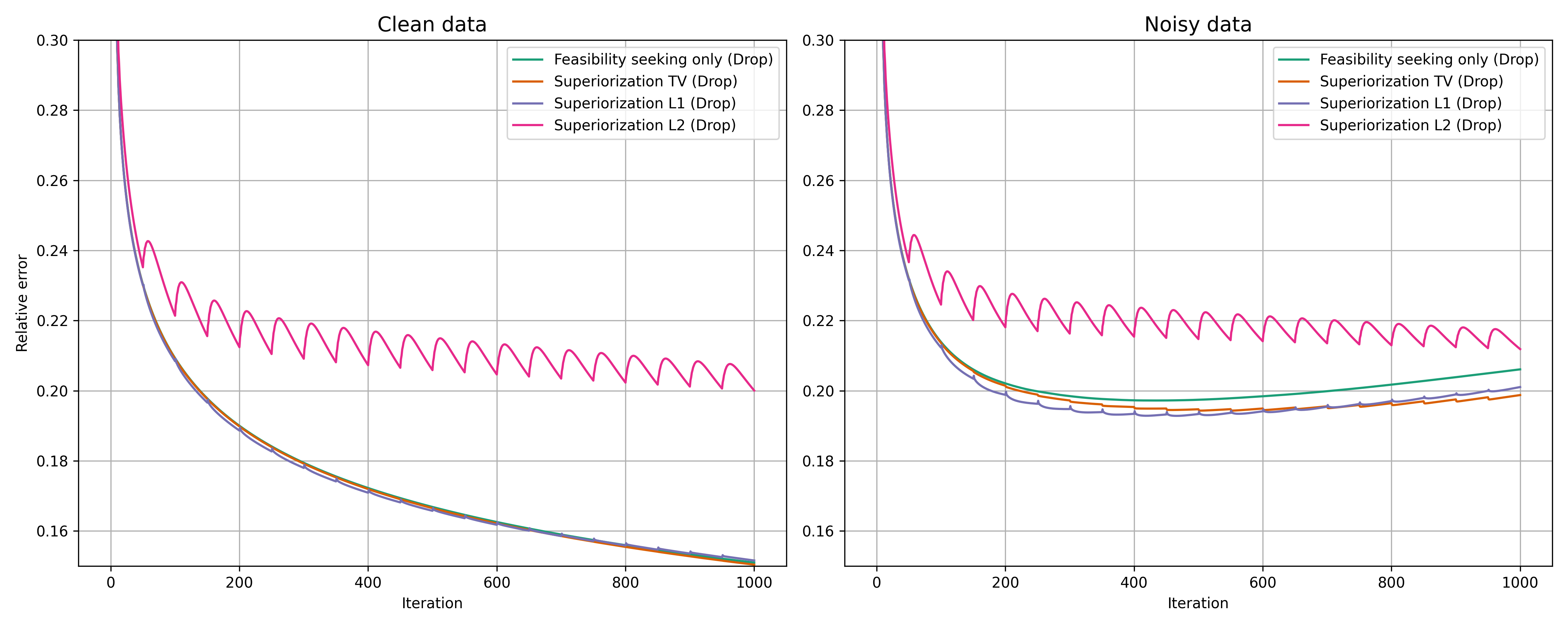}
    \caption{Evolution of the relative errors for the different perturbation strategies}
    \label{fig:seismic_errors}
\end{figure}

\autoref{fig:seismic_errors} shows the evolution of relative error against the number of iterations. In case of the noise-free model ($\mymatrix{A}\myvec{x} = \myvec{b}$) a continuous decrease of the relative error is visible. While differences between feasibility-seeking and the $L1$ and $TV$ perturbations are small, the perturbed versions achieve marginally lower relative errors.

For the noisy data, this improvement for the two perturbation strategies is even more pronounced. 

The $L2$ perturbation on the other hand is not able to achieve this, as the feasibility-seeking on its own outperforms the strategy for both noisy and noise-free data.

\begin{figure}
    \centering
    \includegraphics[width = \textwidth]{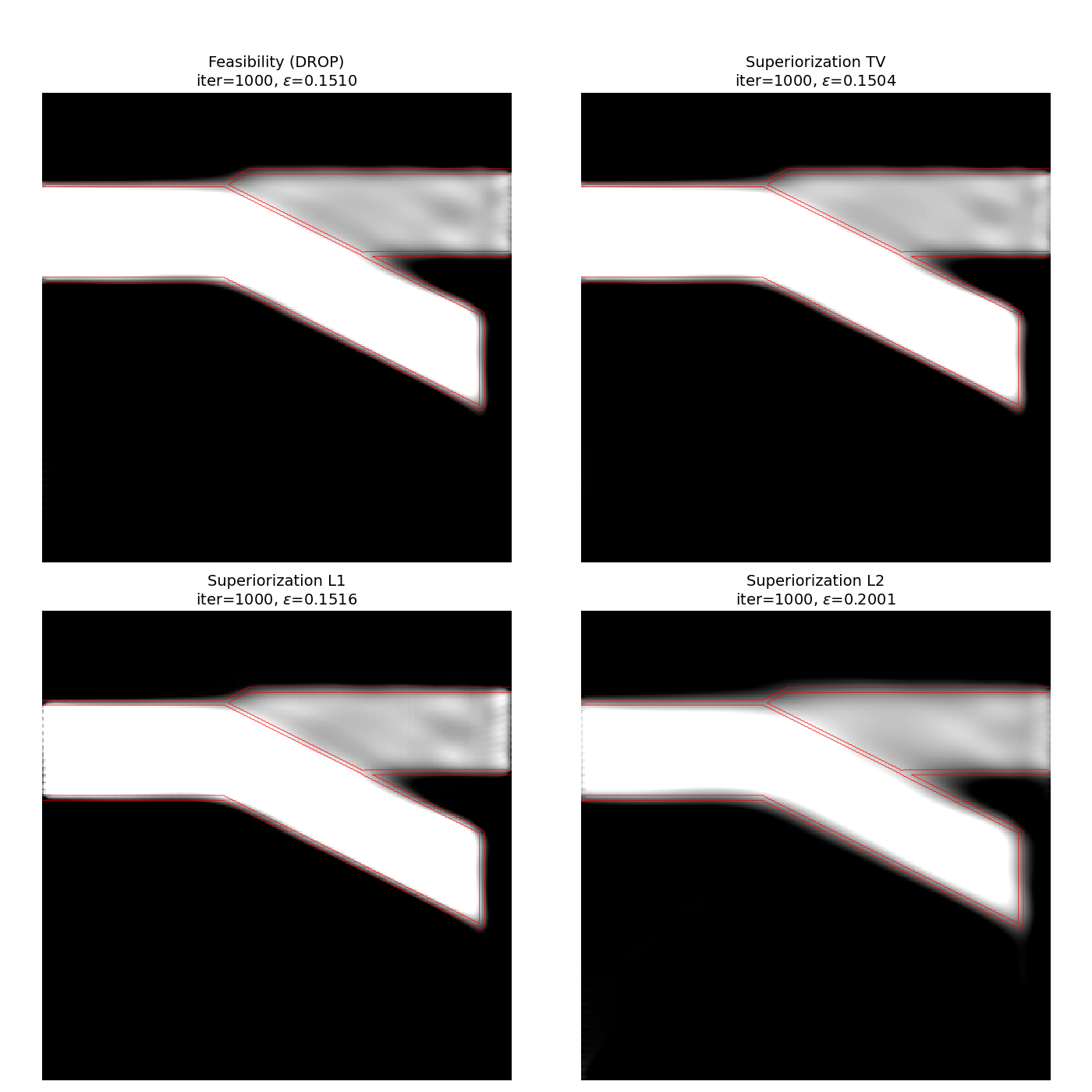}
    \caption{Seismic reconstructions for clean data. Contours of the original tectonic plates are shown in red.}
    \label{fig:seismic_clean}
\end{figure}

\begin{figure}
    \centering
    \includegraphics[width = \textwidth]{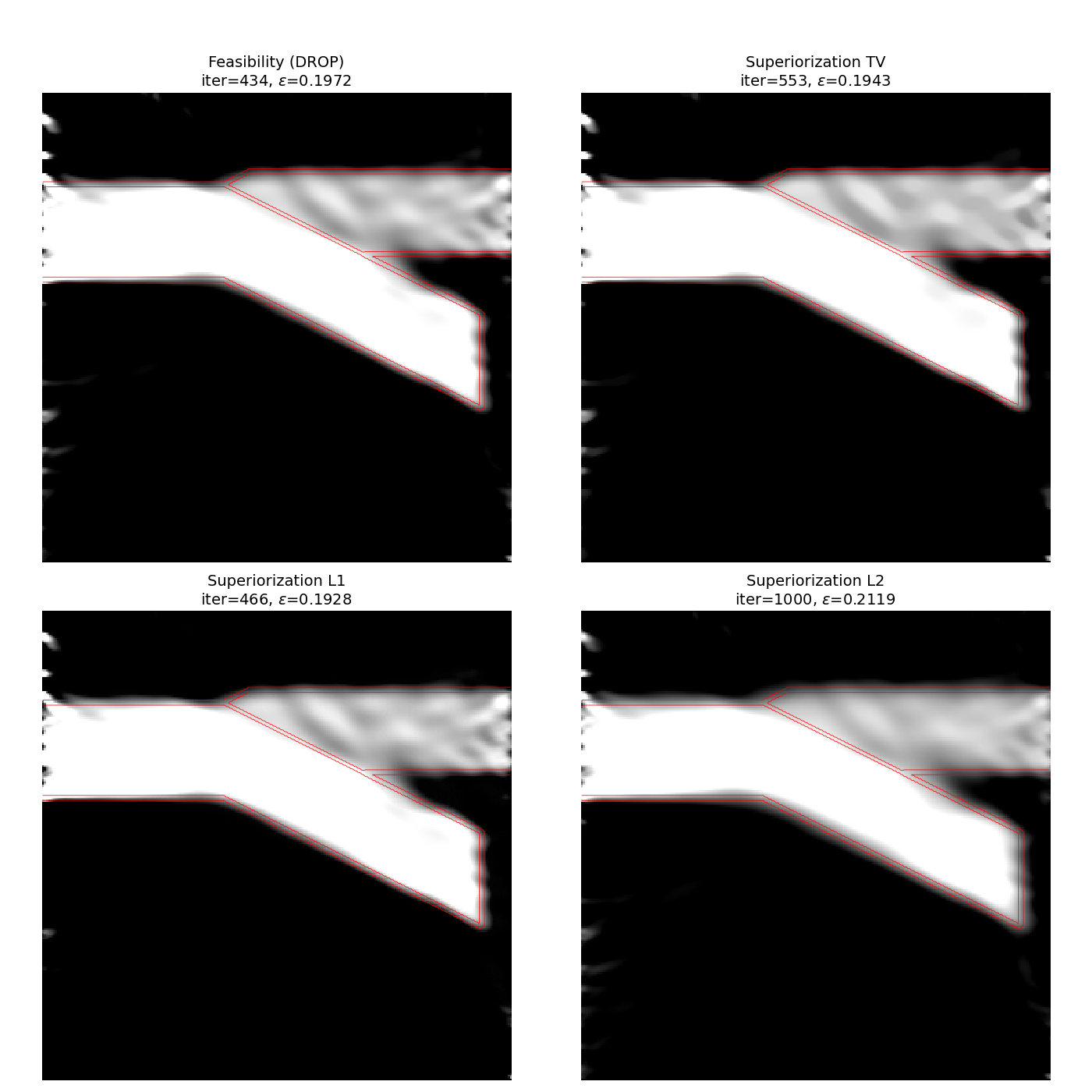}
    \caption{Seismic reconstructions for noisy data. Contours of the original image are shown in red.}
    \label{fig:seismic_noisy}
\end{figure}

However, as shown in \autoref{fig:seismic_clean} and \autoref{fig:seismic_noisy}, all models reconstructed the underlying structures relatively well, although some variations in the right tectonic plates are visible (especially for the noisy data).

\subsection{Application 2: CT Image reconstruction}

\autoref{tab:runtime_ct} shows the runtime for the different feasibility-seeking algorithms and their superiorized versions. The expected increase in runtime was observed when moving from feasibility-seeking to superiorization, due to the added perturbation phase's operations. 
For the simultaneous variants (EMR, EL, CG), both perturbed and unperturbed models took only a few seconds to run on the GPU and the increase due to the perturbations is on the order of 1.5-2. 
When running the same algorithms on the CPU, the runtime increased significantly with the feasibility-seeking algorithms taking about 20 times as long and their superiorized version about 13-20 times as long.
A similar increase could be seen when running on the GPU, albeit the magnitude was lower by a factor of $\sim$ 1.2-1.5.
In contrast, the sequential method's runtime on the CPU heavily exceeded that of the simultaneous variants. Contrary to the simultaneous variants, the runtime increase because of the perturbations can be neglected due to the long runtime of a single iteration of the feasibility-seeking algorithm.

The table also shows the achieved relative errors with and without the early stopping criterion with only minor deviations when stopping based on the variance measure.

\begin{table}[ht]
    \centering
    \resizebox{\textwidth}{!}{%
    \begin{tabular}{lccccc  |cccccc}
    & & \multicolumn{4}{c|}{300 iterations} & \multicolumn{6}{c}{Variance stopping criterion}\\
    Alg. & & t$_{\text{feas}}$\,[s] & $\epsilon_{\min}$ & t$_{\text{sup}}$\,[s] & $\epsilon_{\min}$ 
              & n$_{\text{feas}}$ & t$_{\text{feas}}$\,[s] & $\epsilon_{stop}$ 
              & n$_{\text{sup}}$ & t$_{\text{sup}}$\,[s] & $\epsilon_{stop}$\\ \hline
    ART & CPU & 1447 & $1.0$ & 1578 & $1.0$ & 147 & 697 & $14.569$ & 144 & 753 & $11.392$\\ [0.5em]

    \multirow{ 2}{*}{EMR} & GPU &  10.0 & \multirow{ 2}{*}{0.093} & 17.3 & \multirow{ 2}{*}{0.066} & \multirow{ 2}{*}{109} & 3.7 & \multirow{ 2}{*}{0.110} & \multirow{ 2}{*}{109} & 6.7 & \multirow{ 2}{*}{0.075}\\
    
     & CPU & 235 &  & 348 &  &  & 86 & & & 135 & \\ [0.5em]

    \multirow{ 2}{*}{EL}&  GPU & 7.2 & \multirow{ 2}{*}{0.321} & 15.1 & \multirow{ 2}{*}{0.316} & $\diagup$ & 7.6 & \multirow{ 2}{*}{0.689} & $\diagup$ & 15.0 & \multirow{ 2}{*}{0.661}\\
    
    &  CPU & 182  & & 277 &  & $\diagup$ & 202 &  & $\diagup$ & 244 & \\ [0.5em]

    \multirow{ 2}{*}{CG} & GPU & 9.4 &\multirow{ 2}{*}{0.090} & 15.5 & \multirow{ 2}{*}{0.068} & \multirow{ 2}{*}{115} & 3.7 & \multirow{ 2}{*}{0.861} & \multirow{ 2}{*}{114} & 6.3 & \multirow{ 2}{*}{0.069} \\
    
    & CPU & 229 &  & 286 &  &  & 83 & &  & 111 & \\

    \end{tabular}%
    }
    \caption[Runtime CT-reconstruction]{Results of the CT reconstruction. For each feasibility-seeking algorithm and its superiorized version, the table reports runtimes and relative errors under two settings. The left block terminates after 300 iterations, with $\epsilon_{\min}$ the minimal relative error attained over those iterations. The right block uses the variance-based stopping criterion: n$_{\text{feas}}$ and n$_{\text{sup}}$ are the iterations at which the criterion was met for feasibility-seeking and superiorization, respectively, and $\epsilon_{stop}$ is the relative error at that iteration. t$_{\text{feas}}$ and t$_{\text{sup}}$ denote the runtimes (in seconds) of feasibility-seeking alone and of superiorization.
    }
    \label{tab:runtime_ct}
\end{table}

\begin{figure}
    \centering
    \includegraphics[width = \textwidth]{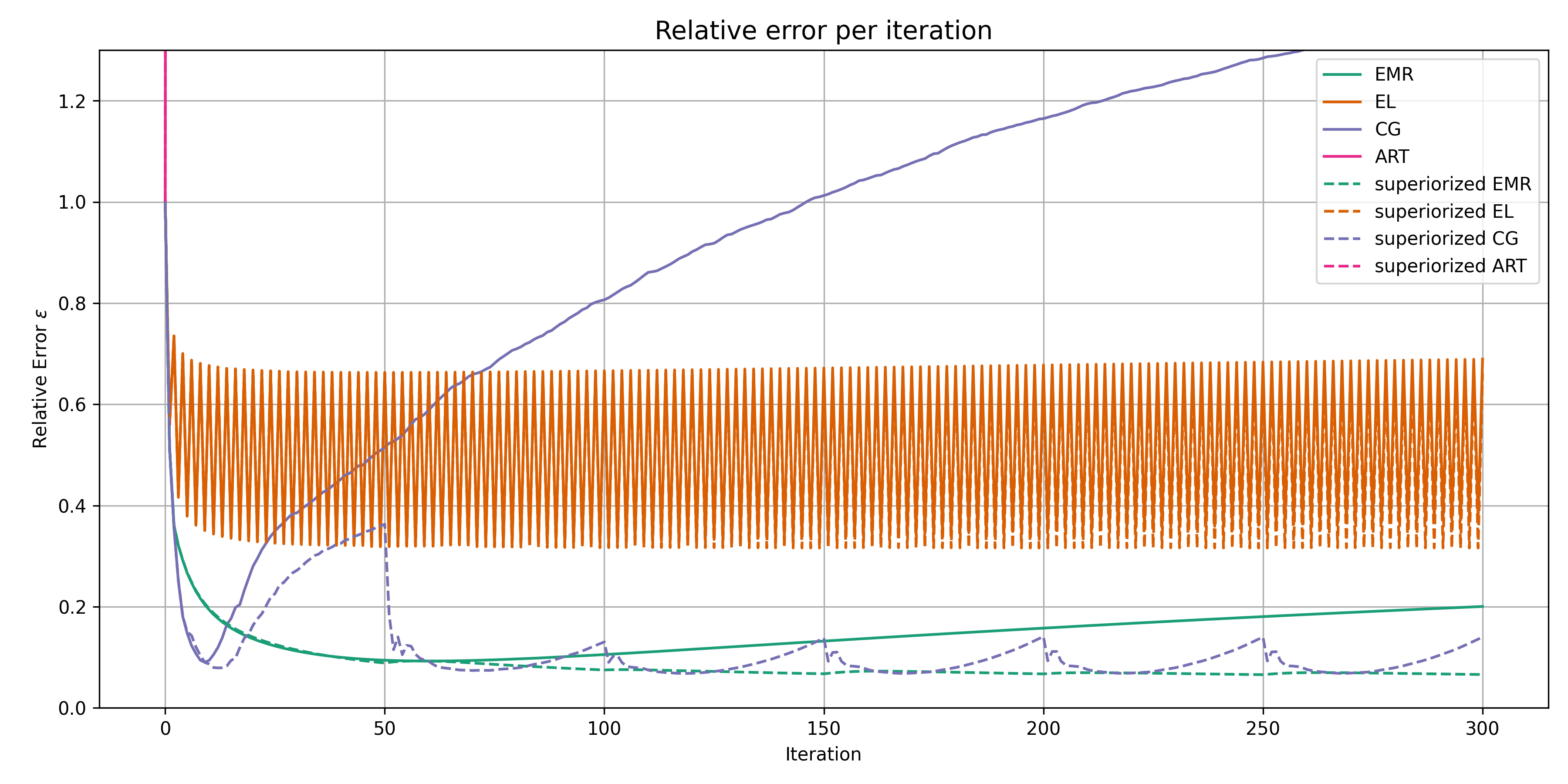}
    \caption{Evolution of relative error for different perturbation strategies. For each algorithm curves for unperturbed and perturbed algorithms are shown. The iterates of the sequential algorithm immediately reach relative errors out of the bounds of the plot and are therefore only visible in the first iteration.}
    \label{fig:rel_error_ct_cropped}
\end{figure}

Furthermore, \autoref{fig:rel_error_ct_cropped} shows the convergence behavior of the tested algorithms. The sequential algorithm was not able to achieve reasonable CT reconstruction quality and led to an increase in relative error. Similarly, the EL algorithm initially decreased in relative error but then started fluctuating for both the perturbed and unperturbed case. This is partly expected, as neither EL nor sequential Kaczmarz is guaranteed to converge on inconsistent feasibility problems ($C = \cap_i C_i = \emptyset$). CG and EMR on the other hand were able to achieve the lowest relative errors, which are further decreased when using their superiorized versions.

\begin{figure}
    \centering
    \includegraphics[width = \textwidth]{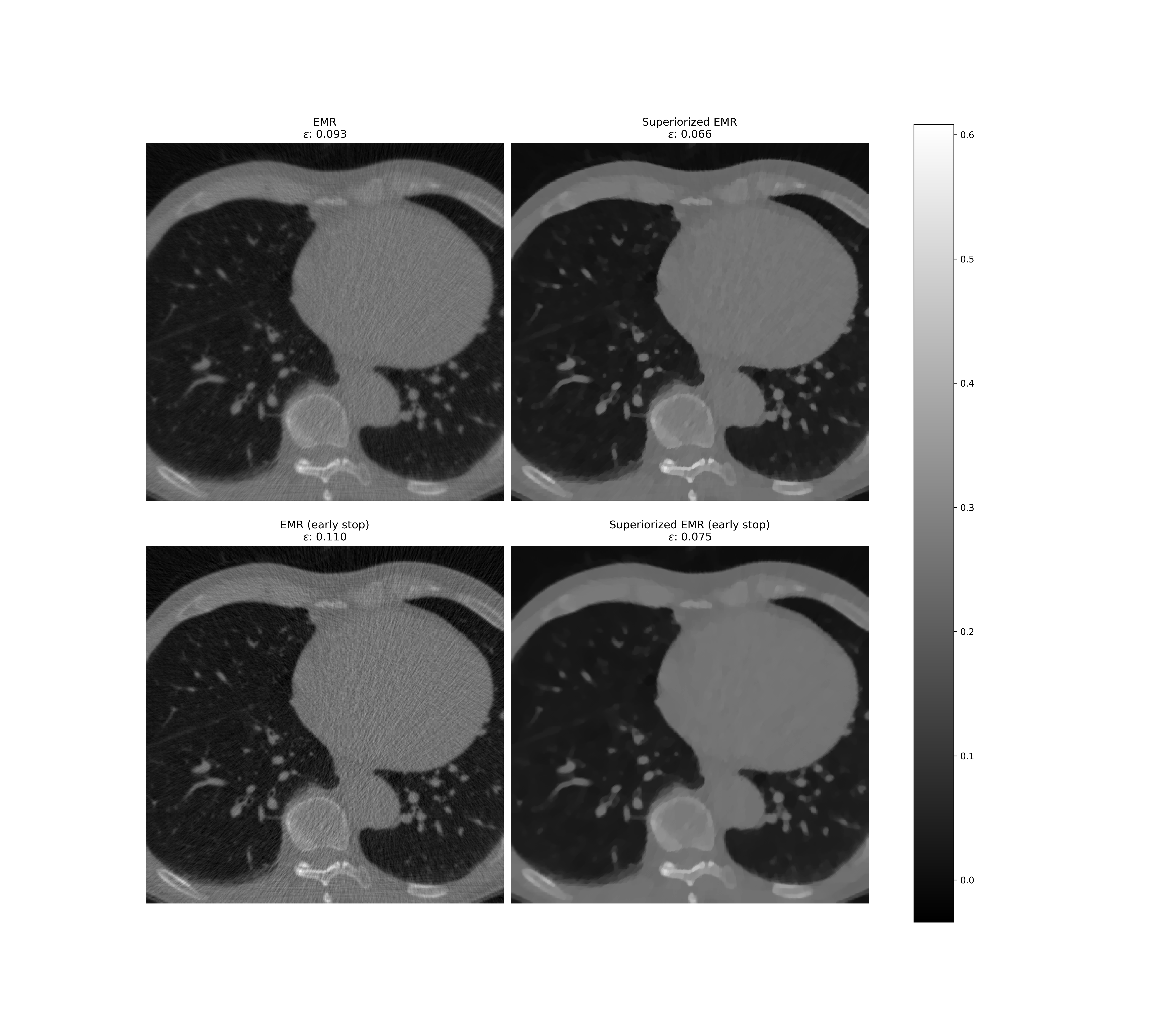}
    \caption{The ground truth CT, the  best reconstruction via feasibility-seeking only and its superiorized version for the error minimizing Landweber method (EMR) are depicted.}
    \label{fig:ct_rec}
\end{figure}

\begin{figure}
    \centering
    \includegraphics[width = \textwidth]{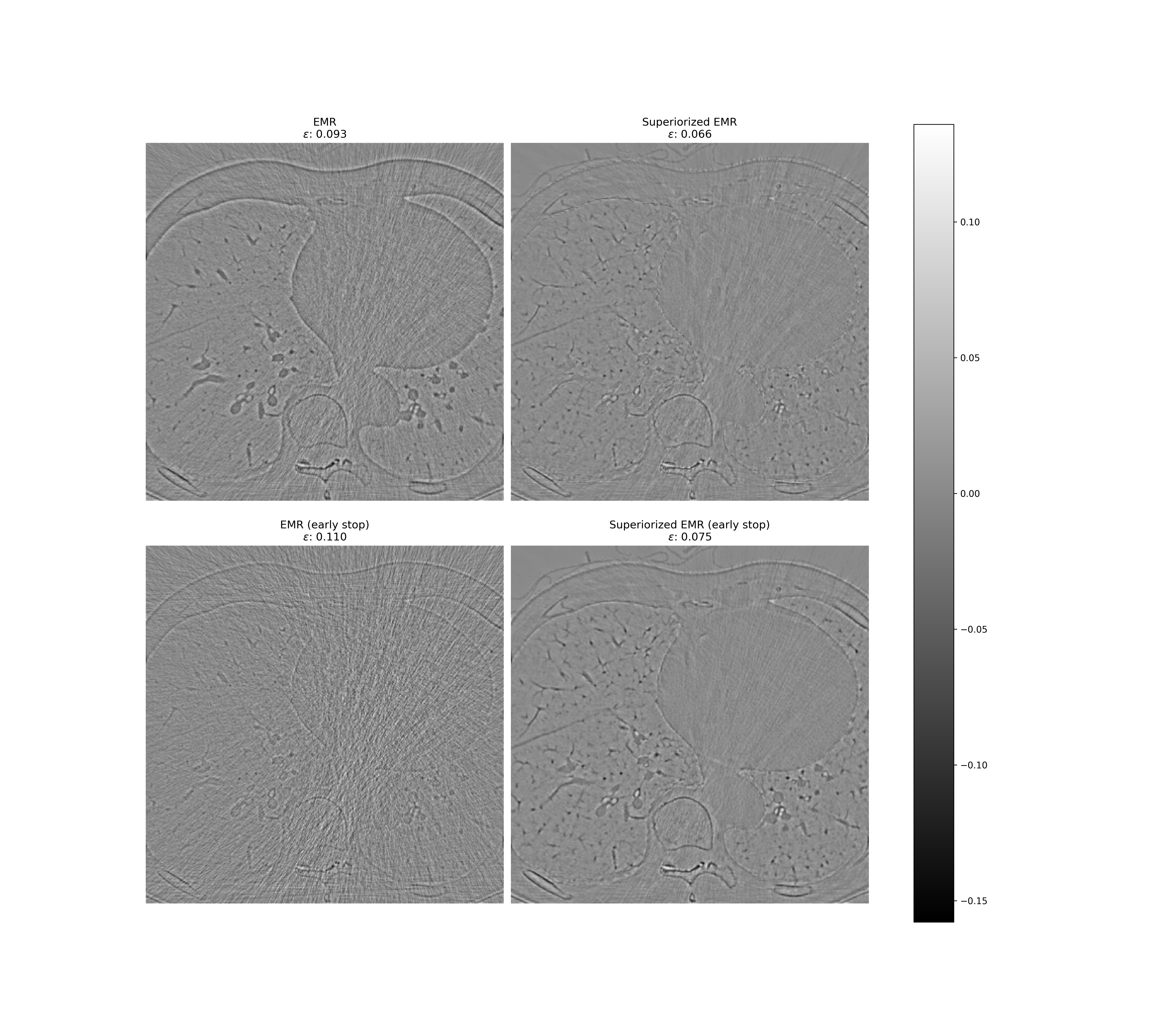}
    \caption{Difference to the true solution for the reconstructions by feasibility-seeking and by superiorization, respectively.}
    \label{fig:ct_rec_diff}
\end{figure}

Qualitative analysis of the reconstructed image slices in \autoref{fig:ct_rec} and \autoref{fig:ct_rec_diff} reveals further details. Comparing the superiorized solution of EMR to the solution of the feasibility-seeking algorithm on its own, both variants reconstructed the anatomical features clearly, with only minor quality loss when the variance-based early-stopping criterion was applied (see \autoref{fig:ct_rec}). However the superiorized version reduced the noise in the reconstructed images, underlined by the lower relative errors than its unsuperiorized version. This is highlighted by \autoref{fig:ct_rec_diff} which shows difference plots for the unperturbed and perturbed solution. The superiorized version achieved visually better reconstruction than feasibility-seeking alone.

\begin{figure}
    \centering
    \includegraphics[width = \textwidth]{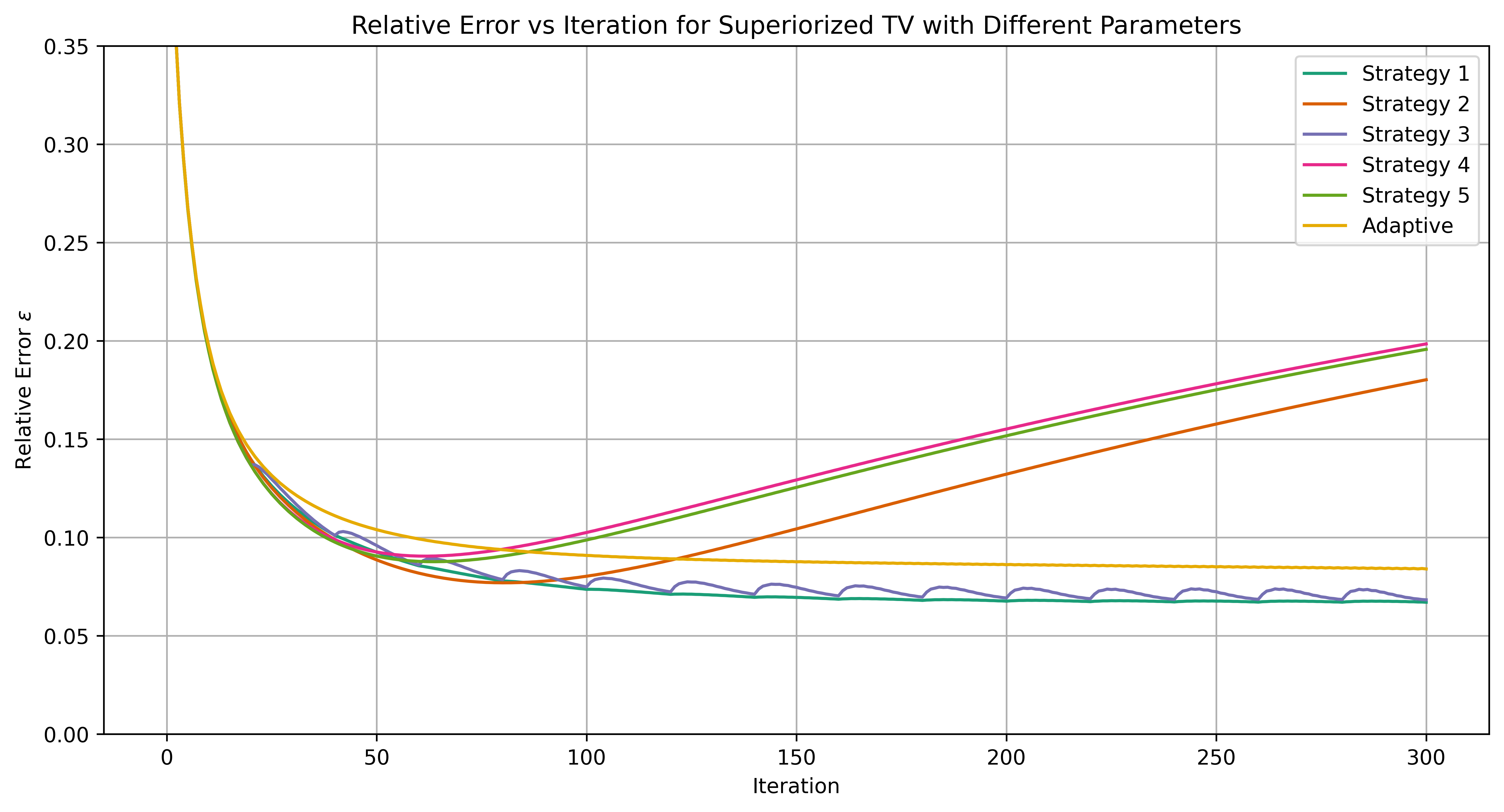}
    \caption{Evolution of the relative error for different perturbation strategies (explanation of the strategies in \autoref{tab:perturbation_parameters}).}
    \label{fig:ct_rel_error_evol_pert}
\end{figure}

\autoref{fig:ct_rel_error_evol_pert} shows the results for the second experiment. The strategies without restart show an L-shape in the error curve, which is not present for the restarted strategies. Furthermore, the adaptive strategy resulted in a lower relative error than the power law strategies without restart, but was outperformed when using a restart.

\subsection{Application 3: Radiotherapy treatment planning}

\begin{figure} 
    \centering
    \includegraphics[width = \textwidth]{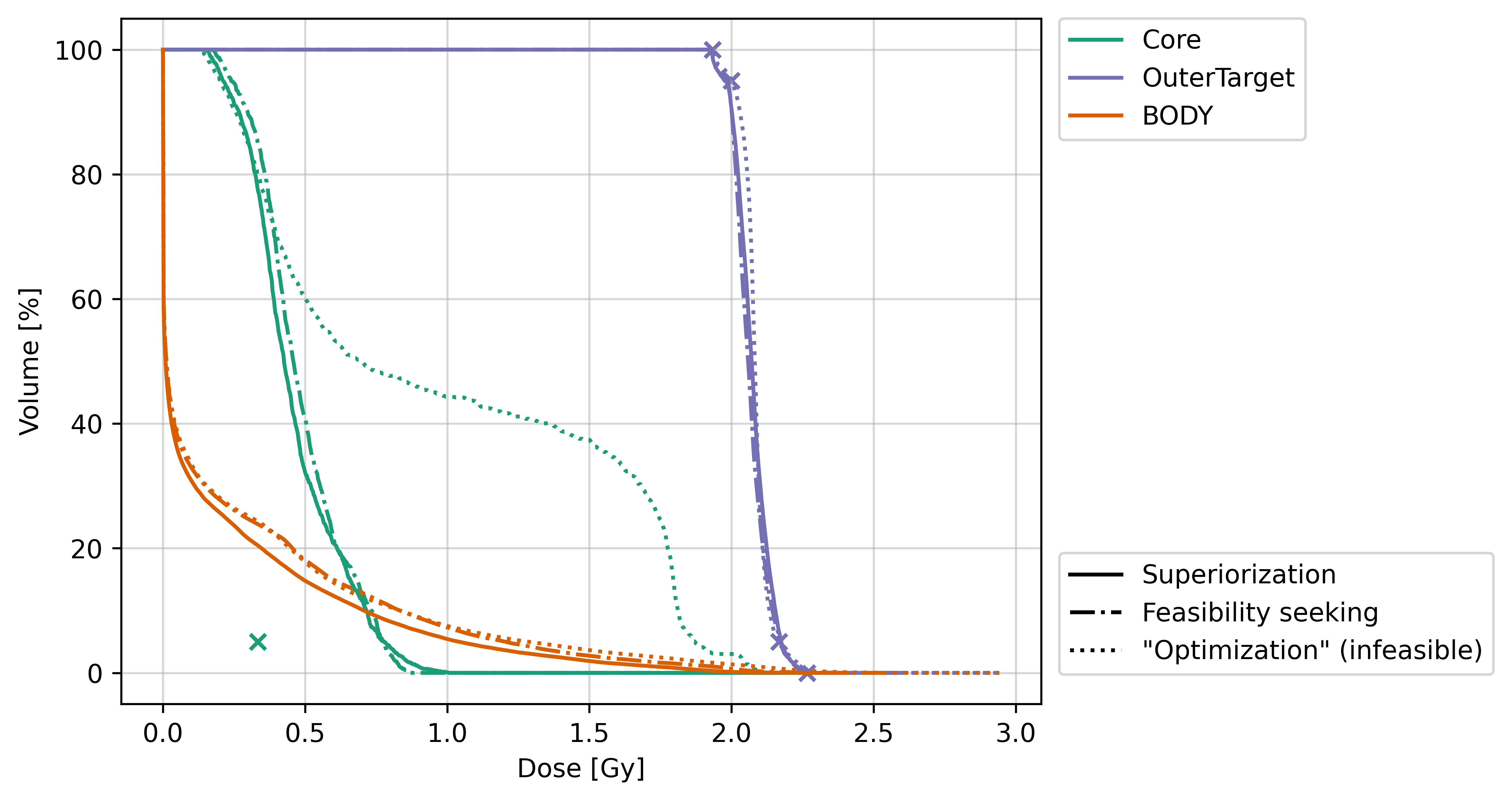}
    \caption{DVH comparison for the three different plan configurations of the horseshoe phantom. The plan label "Optimization" was obtained by the \texttt{IPOPT} solver.}
    \label{fig:dvh_comp}
\end{figure}

\begin{figure}
    \centering
    \includegraphics[width = \textwidth]{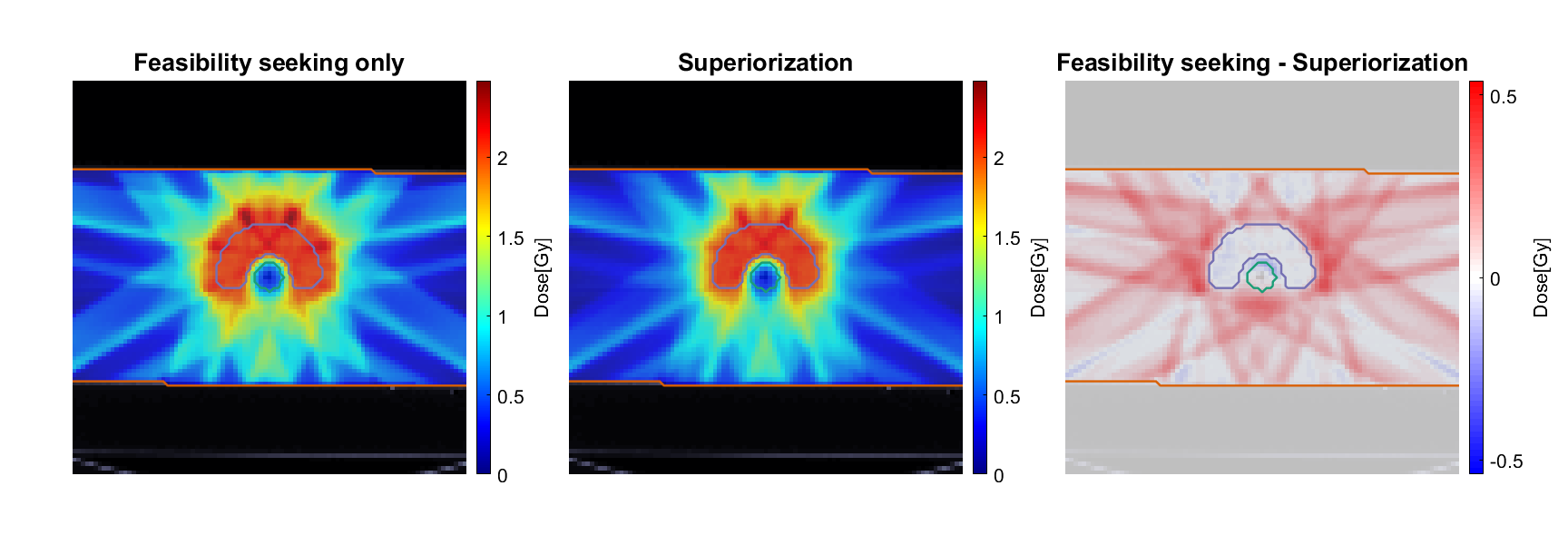}
    \caption{Dose slices for a single fraction computed with feasibility-seeking and superiorization.}
    \label{fig:dose_comp}
\end{figure}

\autoref{fig:dvh_comp} and \autoref{fig:dose_comp} show a dose volume histogram (DVH) and dose distributions for the horseshoe phantom plan, respectively. Both superiorization and feasibility-seeking were started from the same initial point. In contrast to the standard optimizer which struggles with the infeasible constraints employed, feasibility-seeking and superiorization both reach plan configurations that take this into account as well as possible.

Comparing the superiorized version to the plan using feasibility-seeking alone, a reduction of the dose delivered to the surrounding body can be achieved through the perturbation, as seen in the right plot in \autoref{fig:dose_comp}.

The feasibility-seeking algorithm took $105$ seconds to run, while the superiorized version runs for $\sim 300$ seconds, both running on the GPU. Both algorithms were terminated after running 10,000 iterations.

\begin{figure}
    \centering
    \includegraphics[width = \textwidth]{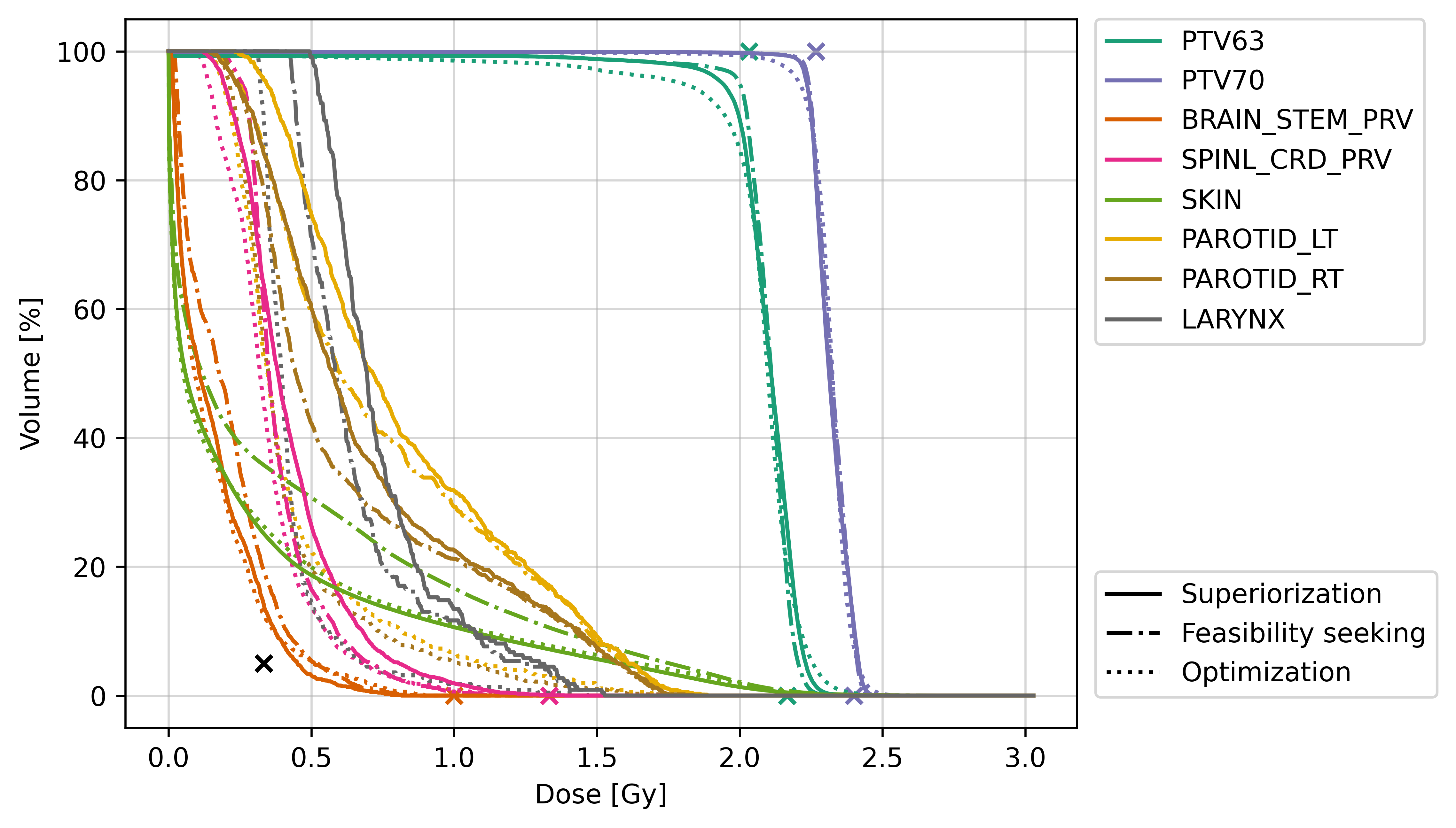}
    \caption{DVH-comparison for the three different plan configurations of the Head and neck patient. The \textit{feasibility-seeking} and \textit{superiorization} plan were calculated using the plan described in \autoref{tab:final_HN_plan}. Furthermore an optimized plan was calculated using only objective functions to show how a clinically relevant plan might look like. The \textit{Superiorization} and \textit{feasibility-seeking} plans show similar qualities, with the \textit{Superiorization} plan additionally reducing the dose in the body, as expected.}
    \label{fig:dvh_comp_HN}
\end{figure}

\begin{figure}
    \centering
    \includegraphics[width = \textwidth]{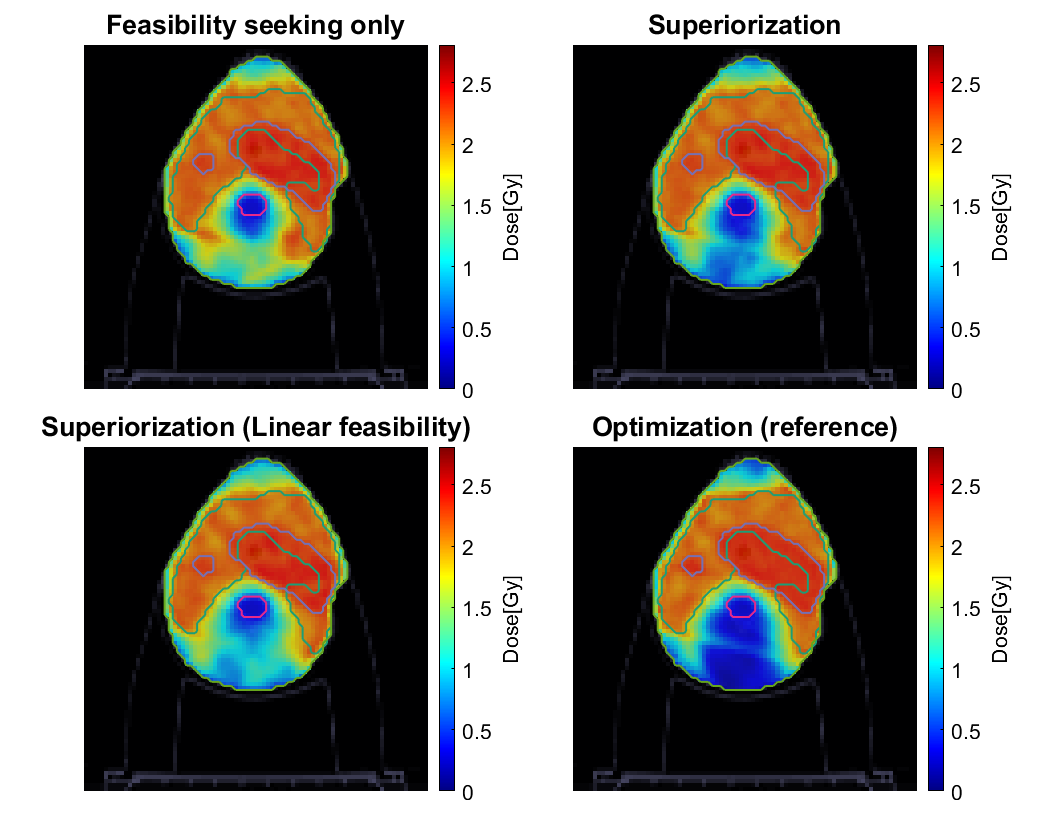}
    \caption{Comparison of different plan slices for the head and neck patient. The top row shows plans based on split-feasibility problems. As visible, the superiorized version reduces the dose in the surrounding body substantially and achieves dosimetric quality comparable to a reference plan calculated with standard optimization (bottom right). The bottom left plan additionally shows a dose distribution achieved through superiorization only employing linear constraints.}
    \label{fig:dose_comp_HN}
\end{figure}

Similar results were achieved on the more demanding head and neck patient.
\autoref{fig:dose_comp_HN} shows dose slices for a head and neck patient. As the two plots in the top row and \autoref{fig:dvh_comp_HN} show, the superiorization algorithm was able to reduce the dose in the body significantly and achieves similar dosimetric quality as the optimized plan.

Running the feasibility-seeking algorithm alone took about 1.75 minutes for 2500 iterations, while the runtime for the superiorized version increased to 4.5 minutes. The plan calculated through CPU-based optimization took about 8.5 minutes. 

While the superiorized plan shows the practical applicability of DVC based constraints, their slow convergence rate reduces their efficiency. Replacing them with linear max-dose constraints achieves a comparable plan quality after 500 iterations (see also \autoref{fig:dose_comp_HN}), while requiring only $50\,s$ of runtime.

\newpage
\section{Discussion}
This work applies the superiorization method to constrained problems in applied physics, which are paired with regularization strategies or secondary objective functions describing soft goals. The method is rigorously implemented in the presented open-source toolbox \texttt{SupPy}, enabling GPU-accelerated feasibility-seeking with adequate perturbation strategies to achieve competitive runtimes for the application of superiorization.

All three demonstrated problems employed superiorization, while focusing on different end goals.

The first experiment tackled seismic image reconstruction on noisy and noise-free data and explored the idea of using superiorization in a similar manner to regularization. Both the L1 and TV strategy were able to outperform their feasibility-seeking algorithm alone, albeit achieving only small improvements. The L2 perturbation on the other hand did not achieve any improvement over the feasibility-seeking strategy on its own.

The second experiment dealt with the reconstruction of CT images and explored the possibility to compare different feasibility-seeking algorithms 
as well as different perturbation strategies. Runtime and convergence analysis were performed for a sequential method run on the CPU as well as three simultaneous methods run on the GPU (EL, EMR and CG). 
Out of all four methods, EMR and CG performed best, while EL and the sequential Kaczmarz/ART strategy struggled to achieve meaningful reconstructions with and without superiorization. For EMR and CG the feasibility-seeking results were further enhanced through the TV-superiorization, enabling a reduction in reconstruction noise.

Comparing different perturbation strategies for the CT reconstruction showed several notable trends: First, an increased perturbation magnitude leads to improved reconstruction quality. Second, the incorporation of restarts (strategy 1 and strategy 3) yields an enhancement in solution quality and mitigates the L-shape of the error curve, thereby reducing susceptibility to noise overfitting. Third, although the adaptive perturbation approach requires fewer input parameters, it is capable of outperforming several power-law–based strategies; nevertheless, power-law approaches with high values of $\alpha$ and possible restarts can achieve superior results.

The possibility of direct comparisons with different algorithms or perturbation strategies, as shown in these experiments is supported by the implementation design of \texttt{SupPy}, providing interfaces for projection algorithms as well as perturbation strategies.

In the final experiment an application to radiotherapy was explored for inconsistent constraints. This could provide an advantage over standard optimization algorithms, as inconsistent problems pose a challenge for optimization algorithms.
Furthermore, the option to use a split-feasibility-seeking problem was demonstrated with dose volume constraints shown to be a viable, albeit slow option for volumetric goals.

Runtime of the used algorithms depends on their respective convergence rate, problem size and stopping criterion. 
For the CT reconstruction the same algorithms were evaluated on the CPU and GPU, with the latter providing a significant speedup due to the faster matrix operations.

Moving from feasibility-seeking to superiorization leads to an expected increase in runtime in all cases, whose amount depends on the computational complexity added by the perturbations. For the simultaneous CT reconstruction the perturbations approximately double the runtime, whereas it takes a multiple of the feasibility-seeking time for both radiotherapy treatment planning cases. This can be explained by looking at the sizes of the matrices employed. For the CT reconstruction the same underlying matrix is used for the calculation of the feasibility-seeking algorithm and perturbation steps. For the investigated radiotherapy plans on the other hand fewer organs are considered for the feasibility-seeking algorithm  than for the perturbations, leading to a larger dose-influence matrix $\mymatrix{A}$ for the latter and therefore more computationally expensive operations.

This work and \texttt{SupPy} thereby serve as an entry point for future application of the superiorization method in physical problems, where secondary goals, like for example noise, are not required to be optimized to absolute minima.

While we report the current state of the toolbox, it is still actively improved upon and extended with new features. Of particular interest may be the addition of dedicated CUDA kernels for implementation of various algorithms, perturbation strategies that do not rely on subgradients, as well as matrix free options for linear algorithms.

\section{Conclusion}
This work demonstrated multiple potential applications for superiorization in applied computational physics with competitive solutions and runtimes. Rigorous implementation of feasibility-seeking algorithms and corresponding perturbation strategies in an open-source toolbox \texttt{SupPy} provides the research community with a platform to explore further applications without re-implementing the underlying algorithmic architecture.

\section*{Declaration of generative AI and AI-assisted technologies in the manuscript preparation process
}
During the preparation of this work the authors used Claude Opus 4.7 and Claude Opus 4.8 by Anthropic in order to spell check and improve the consistency and readability of the manuscript. After using this tool/service, the authors reviewed and edited the content as needed and take full responsibility for the content of the published article.

\section*{Acknowledgments}
This work is supported by the Cooperation Program in Cancer Research
of the German Cancer Research Center (DKFZ) and Israel’s Ministry of Innovation,
Science and Technology (MOST).

\section*{Conflict of Interest}

The authors have declared no conflict of interest.

\newpage
\appendix
\section{Used Methods}
\begin{table}[ht]
    \centering
     \resizebox{\textwidth}{!}{
    \begin{tabular}{cc}
        Method &  Update Step\\
        \hline
        DROP \cite{Censor2008} & $\myvec{x}^{k+1} = \myvec{x}^{k} -  \mymatrix{Z}\mymatrix{A}^{T} \mymatrix{M}(\mymatrix{A}\myvec{x}^k - \myvec{b})$ \\
        \hline
        Sequential/Kaczmarz \cite{Kaczmarz1937} &  $\myvec{x}^{k,i} = \myvec{x}^{k,i-1}  - \frac{\langle \myvec{a}^i,\myvec{x}^{k,i-1} \rangle -b_i}{||\myvec{a}^i||^2}\myvec{a}^i \forall C_i$\\
        Extrapolated Landweber (EL) \cite{Cegielski2013,Cegielski2025} & $\myvec{x}^{k+1} = \myvec{x}^k - \frac{(\mymatrix{A} x^k - \myvec{b})^{T} \mymatrix{D}(\mymatrix{A}\myvec{x}^k - \myvec{b}) }{ \left\lVert \mymatrix{A}^{T} \mymatrix{D}(\mymatrix{A}\myvec{x}^k - \myvec{b}) \right\rVert^{2}} \mymatrix{A}^{T} \mymatrix{D}(\mymatrix{A}\myvec{x}^k - \myvec{b}) $\\
        Landweber with error minimization (EMR) \cite{Nikazad2017}& $\myvec{x}^k - \frac{\left\lVert \mymatrix{A}^{T} \mymatrix{M}(\mymatrix{A}\myvec{x}^k - \myvec{b}) \right\rVert^{2} }{ \left\lVert \mymatrix{M}^{1/2} \mymatrix{A} \mymatrix{A}^{T} \mymatrix{M}(\mymatrix{A}\myvec{x}^k - \myvec{b}) \right\rVert^{2}} \mymatrix{A}^{T} \mymatrix{M}(\mymatrix{A}\myvec{x}^k - \myvec{b}) $\\
        Conjugate Gradient & see Zibetti et al., Algorithm 9 \cite{Zibetti2018}

    \end{tabular}
    }
    \caption{Algorithms used in the seismic tomography and CT reconstruction examples. 
    In the case of the sequential algorithm, one iteration sweeps over all constraints $C_i$, while the other four handle the constraints simultaneously. $D,M$ and $Z$ are diagonal matrices: $D=diag(w_1/||\myvec{a}^1||^2,\dots,w_n/||\myvec{a}^n||^2)$), $M = diag(w_1,\dots,w_n)$ and $Z = diag\Big{(}\text{nnz}(A_{:,1})^{-1},\dots, \text{nnz}(A_{:,m})^{-1}\Big{)}$, where each entry corresponds to the number of nonzero elements in the respective column. In both cases $\sum_i w_i = 1$ holds.}
    \label{tab:ct_algs}
\end{table}

\begin{table}[ht]
    \centering
    \begin{tabular}{cc}
        Name &  Mathematical expression\\ \hline
        Mean Dose ($f_{\text{Mean}})$ & $\frac{1}{N_S} \sum_i d_i$\\
        Squared Deviation ($f_{\text{SqDev}})$ & $\frac{1}{N_S} \sum_i(d_i - d_{ref})^2$\\
        Squared Overdosing ($f_{\text{SqOD}}$) & $\frac{1}{N_S} \sum_i(d_i - d_{ref})_{\text{+}}^2$\\
    \end{tabular}
    \caption{Different objective functions used in optimization for radiotherapy treatment planning. In all cases the sum goes over the dose $d_i$ in all $N_S$ voxels in the structure and $d_{ref}$ is the reference value for the dose $d_i$. For the \textit{Squared} \textit{Overdosing} function only dose values above the reference dose are taken into account for the sum.} 
    \label{tab:dose_objectives}
\end{table}
\bibliographystyle{elsarticle-num-names} 
\bibliography{Superiorization_bibtex_elsevier}

\end{document}